\documentclass[11pt]{article}
\usepackage{jheppub}

\usepackage{slashed}
\usepackage{graphicx}
\usepackage{subfigure}
\usepackage{dcolumn}
\usepackage{bm}
\usepackage{amsfonts}
\usepackage{xcolor}

\def \Lag{\mathcal L}
\def \Act{\mathcal A}
\newcommand{\bwt}{\begin{widetext}}
\newcommand{\ewt}{\end{widetext}}
\newcommand{\newc}{\newcommand}
\newc{\beq}{\begin{equation}}
\newc{\eeq}{\end{equation}}
\newc{\beqa}{\begin{eqnarray}}
\newc{\eeqa}{\end{eqnarray}}
\newc{\nonr}{\nonumber}
\newc{\bi}{\begin{itemize}}
\newc{\ei}{\end{itemize}}
\newc{\ra}{\rightarrow}
\newc{\la}{\leftarrow}
\newc{\lra}{\longrightarrow}
\newc{\lla}{\longleftarrow}
\newc{\Lra}{\Longrightarrow}
\newc{\Lla}{\Longleftarrow}
\newc{\half}{\frac{1}{2}}
\newc{\del}{\delta}
\newc{\Del}{\Delta}
\newc{\eps}{\epsilon}
\newc{\gm}{\gamma}
\newc{\lam}{\lambda}
\newc{\kap}{\kappa}
\newc{\tri}{\triangle}
\newc{\hc}{\dagger}
\newc{\pd}{\partial}
\newc{\wt}{\widetilde}
\newc{\ovl}{\overline}
\newc{\p}{\partial}
\newc{\tchi}{\tilde{\chi}}
\newc{\ds}{\displaystyle}
\newc{\pmt}{\pm\!\pm}
\newc{\PL}{\hat{L}}
\newc{\PR}{\hat{R}}
\newc{\st}{s_\theta}
\newc{\ct}{c_\theta}

\newcommand{\Uel}{U(1)_\ell}
\newc{\msm}{\mathrm{SM}}
\newc{\mtev}{\mathrm{TeV}}
\newc{\Tr}{\mathrm{Tr}}
\newc{\clbl}{\color{blue}}
\newc{\clg}{\color{green}}
\newc{\clr}{\color{red}}

\mathchardef\mhyphen="2D
\newc{\SL}{\not\!\!}

\begin{document}

\title{Neutrino masses and  gauged $U(1)_\ell$ lepton number
}
\author[a,b]{We-Fu Chang}

\author[b]{John N. Ng}

\affiliation[a]{ Department of Physics, National Tsing Hua University, 101 Sec. 2, KuangFu Rd, Hsinchu 300, Taiwan}
\affiliation[b]{TRIUMF Theory Group, 4004 Wesbrook Mall, Vancouver, B.C. V6T2A3, Canada}
\emailAdd{wfchang@phys.nthu.edu.tw}
\emailAdd{misery@triumf.ca}

\abstract{
We investigate the tree-level neutrino mass generation in the gauged $U(1)_\ell$ lepton model recently proposed by us \cite{Chang:2018vdd}. With the addition of one Standard Model(SM) singlet, $\phi_1(Y=0, \ell=1)$, and one SM triplet scalar, $T(Y=-1,\ell=0)$, realistic lepton masses can be accommodated.  The resulting magnitude of neutrino mass is given by  $\sim v_t^3/v_L^2$, where $v_t$ and $v_L$ are the vacuum expectation values of $T$ and $\phi_1$, respectively, and it is automatically of the inverse see-saw type. Since $v_L$ is the lepton number violation scale we take it to be high, i.e. $O \gtrsim  (\mtev)$. Moreover, the induced lepton flavor violating processes and the phenomenology of the peculiar triplet are studied. An interesting bound, $0.1\lesssim v_t\lesssim24.1$ GeV, is obtained when taking into account the neutrino mass generation, $Br(\mu\rightarrow e \gamma)$, and the limits from oblique parameters, $\Delta S$ and $\Delta T$.
Collider phenomenology of the SM triplets is also discussed.

 }
\maketitle

\section{Introduction}

It is now generally accepted that neutrino oscillation data indicate that at least two of
the three active neutrinos have nonvanishing
masses. This cannot be accommodated in the minimal Standard Model (SM)without adding new degrees of freedom
such as two or more SM right-handed neutrinos. However, neutrino masses can be generated by the addition of the Weinberg operator \cite{Wo}, $O_5$. This nonrenormalizable dimension five operator takes the form of $O_5=\frac{y}{\Lambda}\ell_L \ell_LHH$, where $H$ is the SM Higgs field, $\ell_L$ denotes a SM lefthanded lepton doublet, $y$ is a free dimensionless parameter, and $\Lambda$ is an unknown high scale.  After $H$ takes on
a vacuum expectation value $v\simeq 247$ GeV, the electroweak symmetry is spontaneously broken, and  we get a neutrino mass $m_\nu \sim \frac{y v^2}{\Lambda}$. Since data indicate that $m_\nu \lesssim 1$ eV,  depending on the value of $y$, the scale $\Lambda$ can range from 1 to $ 10^{11}$ TeV. This elegant way of generating neutrino masses using only SM fields comes with the price of nonrenomalizability. Furthermore, it reinforces the idea that the SM is an effective theory and the neutrino masses call for its extension.

 Neutrino mass generated from the Weinberg operator is of the Majorana type, and it has lepton number $\ell=2$ provided the conventional lepton number assignments that all SM charged leptons $e,\mu,\tau$ and their associated neutrinos
$\nu_e,\nu_\mu,\nu_\tau$ have $\ell=1$ and all other SM fields carry $\ell=0$ are assumed. Also the anti-leptons
have $\ell=-1$. This is a natural consequence if lepton number is a $\Uel$ symmetry. Thus, the SM is largely invariant under this symmetry with a very small breaking by the Weinberg operator. However, the nature
of this symmetry is unknown. Usually, the total lepton number is taken to be a global symmetry that is
broken at a very high scale $\Lambda \gtrsim 10^{12}$ GeV by two or more SM singlet righthanded neutrinos $N_R$
with Majorana masses of $O(\Lambda)$. Integrating them out gives rise to the Weinberg operator, and this
is the celebrated type I seesaw mechanism \cite{ Glashow:1979nm, GellMann:1980vs,Yanagida:1979as, Mohapatra:1979ia, Schechter:1981cv}. Doing so raises the question of the origin of the Majorana mass bestowed to
$N_R$. One can add a  Majorana mass for $N_R$ by hand. However, our current understanding is that masses of
fermions are generated by the Higgs mechanism.  It is interesting to also to apply this to
 $\Uel$. Doing so will lead to the existence of a Goldstone boson in the physical spectrum which can act as a candidate for dark radiation \cite{CNJW,CN1}.

Moreover, it is phenomenologically and theoretically interesting to investigate the possibility of a gauged $\Uel$
and study the spontaneously broken gauge theory. There are several possibilities. One can gauge the total
lepton number as in \cite{ST}\footnote{For other constructions in conjunction with gauged baryon number  see\cite{otherL_0,otherL_1,otherL_2,otherL_3,otherL_4}.}. One can also gauge a combination of lepton generation number such as
$L_\mu -L_\tau$\cite{He:1990pn,He:1991qd}. In Ref.(\cite{Chang:2018vdd}), hereafter referred to as (I), we gauged each lepton family with the usual lepton
number assignments for them. Of the just mentioned three examples only the second one is anomaly-free with only the
SM fields. Gauging the total lepton will require extra leptons with very exotic lepton charges such
as $\ell=3$ to cancel the anomalies from $\Uel$.  In (I), the extra anomalies cancelations require two extra pairs of vector-like $SU(2)$ doublet leptons with eigenvalues $\ell=1,0$ for each family. We also did not include any
singlet $N_R$ field, and the Weinberg operator is generated radiatively at 1-loop. The principal source of lepton number violation comes from
a SM singlet scalar with $\ell=2$ which picks up a vacuum expectation value.

In this paper, we study a different mechanism of neutrino mass generation in the gauged lepton number scheme
introduced in (I). The extra leptons presented before is sufficient to generate neutrino masses with the aid of a SM triplet scalar $T$ and a SM singlet scalar $\phi_1$.
$T$ has $\ell=0$ whereas $\phi_1$ is given $\ell=1$, with both fields being Higgssed. This naturally leads to
an inverse seesaw mechanism (ISM)\cite{ISS1,ISS2,ISSr} for active neutrino mass. The novel feature here is that we do not add by hand
any SM singlet leptons to implement ISM as is commonly done. The required leptons are dictated by anomaly cancelations.  Details will be given in Sec. 3.
Since the physics involved with the gauge new gauge boson $Z_\ell$ and the extra leptons are the same
as in (I), we will not repeat their phenomenology here. Instead, we focus on neutrino physics and the
phenomenology of $T$.  We find that $T$ has interesting different signatures at high energy colliders from previous studies of $l=2$ Higgs triplets\cite{tri_1,tri_2,tri_3,tri_4,tri_5}, which are commonly employed in the type-II see-saw model\cite{ss2, ss2_0,ss2_1,ss2_2,ss2_3,ss2_4}. For a recent review see \cite{Cai:2017mow}.

We organize the paper as follows. The next section we present our anomalies solution for completeness.
Then we discuss lepton mass generation for one generation to illustrate the physics. This is followed by
a realistic 3-generation study. Sec. 4 gives fits to the neutrino oscillation data. Constraints from
charged lepton flavor changing neutral currents are given in Sec.5. Important electroweak
precision constraints are studied in Sec. 6.  The productions of different new triplet scalars at
the LHC  and CLIC are examined in Sec. 7.  Our conclusions are given in Sec. 8.

\section{$\Uel$ anomalies cancelations and new fields}
We extend the SM gauged group by adding a $\Uel$ and is explicitly given as $G=SU(2)\times U(1)_Y\times\Uel$.
All SM leptons have the conventional value of $\ell=1$. We will concentrate on one family.
This can be trivially extended for all 3 SM families.

The new anomaly coefficients are
 \begin{subequations}
 \label{eq:ac}
 \beqa
 \Act_1([SU(2)]^2\Uel)&=&-1/2\,, \\
 \Act_2([U(1)_Y]^2\Uel)&=&1/2\,, \\
 \Act_3([U(1)_Y[\Uel]^2)&=&0\,, \\
 \Act_4([\Uel]^3)&=&-1\,,\\
 \Act_5(\Uel)&=&-1\,,
 \eeqa
 \end{subequations}
 where $\Act_5$ stands for the lepton-graviton anomaly.
 While new chiral leptons are introduced to cancel Eq.(\ref{eq:ac}), one also needs to make sure that  the SM anomalies of $\Act_6([SU(2)]^2 U(1)_Y)$, $\Act_7([U(1)_Y]^3)$,  and $\Act_8(U(1)_Y)$ are canceled.
  It is easy to check that the new leptons in Table.\ref{tb:lA} cancel the above anomalies.
\begin{table}
\begin{center}
\renewcommand{\arraystretch}{1.30}
\begin{tabular}{|c|c|c|c|}
\hline
Field&$SU(2)$&$\phantom{U}Y\;\;$&$\phantom{U}\ell\;\;$\\ \hline
$\ell_L=\begin{pmatrix} \nu_L\\e_L \end{pmatrix}$ &{\bf{2}}&$-\frac{1}{2}$&$\phantom{-}1$\\ \hline
$e_R$&{\bf{1}}&$-1$&$\phantom{-}1$\\ \hline
$L_{1L}=\begin{pmatrix}N_{1L}\\E_{1L}\end{pmatrix}$&\bf{2}&$-\frac{1}{2}$&$-1$\\ \hline
$E_{1R}$&{\bf{1}}&$-1$&$-1$ \\ \hline
$L_{2R}=\begin{pmatrix}N_{2R}\\E_{2R}\end{pmatrix}$&{\bf{2}}&$-\frac{1}{2}$&$\phantom{-}0 $\\ \hline
$E_{2L}$&{\bf{1}}&$-1$&$\phantom{-}0 $\\ \hline
\end{tabular}
\caption{Lepton fields for anomalies free solution.}
\label{tb:lA}
\end{center}
\end{table}
Since the pair of new leptons are vectorlike, the SM anomalies $\Act_6([SU(2)]^2 U(1)_Y)$, $\Act_7([U(1)_Y]^3)$, and $\Act_8(U(1)_Y)$ cancelations are not affected.

 The minimal set of scalar fields, by utilizing the triplet scalar for neutrino mass generation,  can be obtained by examining the gauge invariant set of
lepton bilinears that can be formed from the above fields. They are given in Table.\ref{TAB:Scalar_content}.
\begin{table}[h!]
\begin{center}
\renewcommand{\arraystretch}{1.30}
\begin{tabular}{|c|c|c|c|}
\hline
Field&$SU(2)$&$\phantom{U}Y\;\;$&$\phantom{U}\ell $ \\ \hline
$H$ &{\bf{2}}&$\phantom{-}\frac{1}{2}$&$\phantom{-}0$ \\ \hline
$T$&{\bf{3}}&$-1$&$\phantom{-}0 $ \\ \hline
$\phi_1$&\bf{1}&$\phantom{-}0$&$\phantom{-}1$ \\ \hline
\hline
\end{tabular}
\caption{Scalar fields content }
\end{center}
\label{TAB:Scalar_content}
\end{table}
where all $H,\phi_1$, and $T$ develop non-zero VEVs.

The Yukawa interactions are
\beqa
{\Lag}_Y&=&
f_1 \overline{l_L} L_{2R} \phi_1  + f_2\overline{e_R} E_{2L} \phi_1 + f_3\overline{L_{1L}} L_{2R} \phi_1^* + f_4 \overline{E_{1R}} E_{2L} \phi_1^*  \nonr\\
&+&h_1 \overline{l}_L e_R H + h_2 \overline{L_{1L}} E_{1R} H + h_3\overline{L_{2R}}  E_{2L} H \nonr\\
&+& y_1 \overline{l^c_L}T^\dagger L_{1L} + \frac{y_2}{2}\overline{L^c_{2R}}T^\dagger L_{2R}  + h.c.
\label{eq:all_Yukawa}
\eeqa
where all the generation indices are suppressed.
The full gauge invariant and renormalizable  scalar potential reads,
\beqa
V&=&
-\mu_H^2 H^\dagger H  +\lambda_H (H^\dagger H)^2
-\mu_L^2 |\phi_1|^2  + \lambda_L |\phi_1|^4 \nonr \\
&&- \mu_t^2 \Tr(T^\dagger T) +\lambda_t  [ \Tr(T^\dagger T) ]^2 \nonr \\
&&+\lambda_1 (H^\dagger H) \Tr(T^\dagger T)
+\lambda_2 (H^\dagger H) |\phi_1|^2 +\lambda_3   \Tr(T^\dagger T) |\phi_1|^2
 \nonr \\
&&+\lambda_4\Tr(T^\dagger T T^\dagger T)+\lambda_5 \det T^\dagger T+\lambda_6 H^\dagger T T^\dagger H \nonr\\
&&-\sqrt{2}\kappa H^T(i\tau_2) T(i\tau_2) H
+h.c.
\eeqa
where we have used the bi-doublet form for $T$ as below\footnote{If a doublet, $D$, transforms under $SU(2)$ as $D\ra U_2 D$, then $T\ra U_2 T U_2^T$.}
\beq
T=  \left( \begin{array}{cc}
           T_0 & \frac{1}{\sqrt{2}}T_- \\ \frac{1}{\sqrt{2}}T_- & T_{--}
          \end{array}
\right)\,.
\eeq

The following conditions must hold ($\lam_{4t}=\lambda_4\!+\!\lambda_t$)
\beqa
\lambda_H, \lambda_L,\lambda_{4t} >0\,,\;
\lambda_1>-2\sqrt{\lambda_H \lambda_{4t}}\,,\;
\lambda_2 >-2\sqrt{\lambda_H \lambda_L}\,,\;
\lambda_3>-2\sqrt{\lambda_L \lambda_{4t}}\;,
\eeqa
so as to ensure that the potential is bounded from below.

After SSB,
\beq
\langle H \rangle= \frac{v}{\sqrt{2}} \left( \begin{array}{c}  0\\  1\end{array}\right),
\langle \phi_1 \rangle=\frac{v_L}{\sqrt{2}},
\langle T \rangle=  \frac{v_t}{\sqrt{2}}\left( \begin{array}{cc} 1 &0\\ 0&0 \end{array}\right)\,,
\eeq
and the fields can be expanded around their VEVs as
\beq
H= \left( \begin{array}{c}  H_+\\ { v+ \Re H_0 +i \Im H_0 \over\sqrt{2}}\end{array}\right)\,,\,
\phi_1= { v_L+ \Re \Phi +i \Im \Phi \over\sqrt{2}}\,,\,
T=\left( \begin{array}{cc}  { v_t+ \Re T_0 +i \Im T_0 \over\sqrt{2}} & \frac{1}{\sqrt{2}}T_-\\ \frac{1}{\sqrt{2}}T_-&  T_{--}\end{array}\right)\,.
\eeq
And the minimal condition for the scalar potential become
\beqa
&&v\left(-\mu_H^2 +\lambda_H v^2 +\frac{\lambda_1}{2}v_t^2+\frac{\lambda_2}{2}v_L^2 -\kappa v_t \right) =0\,,\\
&&v_L\left(-\mu_L^2 +\lambda_L v_L^2 +\frac{\lambda_2}{2}v^2+\frac{\lambda_3}{2}v_t^2 \right ) =0\,,\\
&&v_t\left(-\mu_t^2 +\lambda_t v_t^2 +\frac{\lambda_1}{2}v^2+\frac{\lambda_3}{2}v_L^2+\lambda_4 v_t^2\right) -\frac{1}{2}\kappa v^2 =0\,.
\eeqa

Note that $\lambda_{5,6}$ do not come into play here.
From the above equations, the tree-level mass squared for $T_-$ and $\Re T_0$ are
\beq
M^2_{T_-}=\frac{\kappa v^2 }{2 v_t} +\frac{\lambda_6}{4}v^2\,,\;
M^2_{\Re T_0}=\frac{\kappa v^2 }{2 v_t}+\frac{1}{2}(\mu_t^2+\lambda_t v_t^2 +\lambda_4 v_t^2)\,.
\label{eq:TL_triplet_mass}
\eeq
From phenomenology we expect that $v_t\ll v$ (see Sec.8) and  before scalar mixings considerations we have
\beq
M_{T_-} \simeq  \left(\frac{\kappa}{v_t}\right)^{\frac{1}{2}} \frac{v}{\sqrt{2}}\,,\;
M_{\Re T_0} \simeq M_{T_-} \sqrt{1+\frac{v_t}{\kappa}\frac{\mu_t^2}{v^2}}
\eeq
which are above a TeV if $\kappa$ takes a phenomenologically interesting value around the electroweak scale.
However, in general, $\kappa$ is a free parameter.

The scalar potential after SSB gives a small mass splitting between $T^-$ and $T^{--}$.
The mass squared difference can be worked out to be
\beq
M^2_{T_{--}}-M^2_{T_-}= v^2 \frac{\lambda_6}{4} -v_t^2 \left(\lambda_4-\frac{\lambda_5}{2}\right)\,,
\eeq
which is  $\sim {\mathcal{O}}(v^2)$ provided $\lam_6$ is not much smaller than $\lam_{4,5}$.
Therefore, it is a good approximation to assume that $T^-$ and $T^{--}$ are degenerate.
However, we should keep in mind the mass splitting could be about the Fermi scale.

Similarly, ignoring the contribution from $v_t$, we have
\beqa
-\mu_H^2+\lambda_H v^2 +\frac{1}{2} \lambda_2 v_L^2 \simeq 0\,,\\
-\mu_L^2+\lambda_L v_L^2 +\frac{1}{2} \lambda_2 v^2 \simeq 0\,.
\eeqa
Since we expect that $v_L \gg v$, i.e. lepton symmetry breaking to be above the Fermi scale, we
obtain
\beq
 v_L \simeq \sqrt{ \frac{\mu_L^2}{\lambda_L}}\,,\,\, M_{\phi_1}\simeq \sqrt{2}\mu_L\,,\,\,
 v^2 \simeq \frac{1}{\lambda_H} \left(\mu_H^2 -\frac{\lambda_2}{4\lambda_L}M^2_{\phi_1}\right)\,.
 \eeq
Thus,  it is also required to have $|\lambda_2| \ll \lambda_L$. As expected, there will be mixing among the three neutral scalars $\mathbf{H}=(\Re H_0, \Re T_0,\Re \Phi)$. They are related to the  physical states $\mathbf{h}=(h_{SM},t_0,\phi_0)$ via the  usual unitary rotation given by
\beq
\label{eq:Uh}
\mathbf{h} =\mathbf{U}_h\cdot \mathbf{H}\,.
\eeq
Details of this transformation are not
important for this study and we will not present them.

For completeness, we discuss the imaginary parts of the scalar fields.  $\Im \Phi$ is the would-be Goldstone for the gauge boson $Z_\ell$.
Moreover,  the would-be Goldstone bosons eaten by $W^\pm, Z$,  the physical singly charged scalars, $h^\pm$, and the pseudoscalar, $A_0$, can be identified as:
\beqa
G_\pm &=& {v H_\pm -\sqrt{2} v_t T_\pm \over \sqrt{v^2+2v_t^2}}\simeq H_\pm\,,\,\, G^0={v \Im H_0-2v_t \Im T_0 \over \sqrt{v^2+4v_t^2}}\simeq \Im H_0\,,\nonr\\
h_\pm &=& {\sqrt{2} v_t H_\pm + v T_\pm \over \sqrt{v^2+2v_t^2}}\simeq T_\pm\,,\,\, A_0={2v_t \Im H_0+v \Im T_0 \over \sqrt{v^2+4v_t^2}}\simeq \Im T_0\,.
\eeqa
Since $v_t\ll v$  from the electroweak precision studies (see Sec. 8), it is a good approximation to treat $T_\pm$ and $\Im T_0$ as the physical states.
Being the only degree of freedom with two units of electric charge, $T_{\pmt}$ are the physical scalars.

Since the symmetry $G$ forbids $T$ from coupling to  two SM fermions simultaneously, its gauge interactions become
the most relevant for phenomenology. From the $G$-covariant derivative we obtain the Feynman
rules for its triple couplings to gauge bosons, displayed in Table \ref{tab:TTV_vertex}, where $P$ stands for the photon, and all the momenta are incoming.
\begin{table}
\begin{center}
\begin{tabular}{|ccc|ccc|}
  \hline
 $T_0^{*} T_- W^+_\mu $  & : & $i g_2(p_--\bar{p}_0)_\mu$ &  $T_+ T_{--} W^+_\mu $   & : & $i g_2(p_{- -}- p_+)_\mu$ \\
 $ T_+T_0  W^-_\mu $  & : & $i g_2(p_0 -p_+ )_\mu$ &  $T_{+\!+} T_{-} W^-_\mu $   & : & $i g_2(p_- - p_{++})_\mu$ \\
  $ T_+ T_- P_\mu $  & : & $ -i e(p_- -p_+ )_\mu$ &  $T_{+\!+} T_{--} P_\mu $   & : & $ -2i e(p_{--} - p_{++})_\mu$ \\
$ T_+ T_- Z_\mu $  & : & $ i\frac{g_2}{c_W}(s_W^2) (p_- -p_+ )_\mu$ &  $T_{+\!+} T_{--} Z_\mu $   & : & $ i\frac{g_2}{c_W}(-1+2s_W^2)(p_{--} - p_{+\!+})_\mu$ \\
$ T_0^{*}T_0 Z_\mu $  & : & $ i\frac{g_2}{c_W} (p_0 -\bar{p}_0 )_\mu$  &  $ \Re T_0 \Im T_0 Z_\mu$   & : & $\frac{g_2}{c_W} (p_{\tiny \Im T_0} - p_{\tiny \Re T_0} )_\mu$ \\
  \hline
\end{tabular}
\caption{Couplings of gauge bosons to triplet fields}
\label{tab:TTV_vertex}
\end{center}
\end{table}

\section{Lepton masses for 1 generation}
The physics of  how the new leptons affect the SM charged leptons is best seen in the
one family scenario.
In the basis $\{e, E_1, E_2 \}$, the Dirac mass matrix is
\beq
{\cal M}_C =\frac{v_L}{\sqrt{2}} \times \left(  \begin{array}{ccc}
                       h_1 \epsilon_v & 0 & f_1 \\
                       0 & h_2 \epsilon_v & f_3 \\
                       f_2^* & f_4^* & h_3 \epsilon_v \\
                     \end{array}
\right)\,,
\eeq
where $\epsilon_v= \frac{v}{v_L}\ll 1$. In general the electron will mix with $E_{1,2}$ and the mixing depends on $f_1$ and $f_2$.
 In that case, the charged-current interaction of the SM leptons could deviate from the canonical SM $(V-A)$ form due to their mixings with $L_{2R}$ and $E_{2L}$. Moreover, the SM gauge couplings are flavor non-diagonal.
Physically, this mixing must be very small and we can take the limiting case of $f_1=f_2=0$ and eliminate the mixing of the electron with the new charged leptons\footnote{ Theoretically,  these two Yukawa couplings can not be forbidden by any $U(1)$ or $Z_N$ charge assignment in this model.
However, one can obtain the desired Yukawa
hierarchy in the split-fermion model if the 5D wave-functions centers are  in the order
of $e_R, l_L, L_{1L}, L_{2R}; E_{1R};E_{2L}$, where ``,'' and ``;'' mean large and small separations in between two adjacent wave functions, respectively, along the fifth dimension, see \cite{SF,RS1_1,RS1_2} for some other examples of achieving the hierarchical 4D Yukawa.  }.
In general, we can write the physical mass eigenstates $\mathcal{E}_\alpha^\prime=(e,E_-,E_+)$ where $\alpha=1,2,3$ as
\beq
\label{eq:DMe}
\mathcal{E}_{L/R}^\prime=V_{L/R}\cdot \mathcal{E}_{L/R}\,,
\eeq
 where $V_{L/R}$ is the left-handed/right-handed unitary matrix that diagonalizes the charged lepton mass matrix so that $V_L^\dagger\cdot M_C\cdot V_R=\mbox{diag}\{m_e,M^E_-,M^E_+\}$. For the limiting case of $f_1=f_2=0$ and $f=f_3=f_4(1+\delta)$ with $|\delta|\ll 1$, the mass eigenvalues can be worked out  to be
 \beq
 m_e=h_1 \frac{\epsilon_v v_L}{\sqrt{2}}\,,\, M^E_{\pm}\simeq\pm \frac{f_4 v_L}{\sqrt{2}}\left(1+\frac{\delta}{2}\right) + \frac{\epsilon_v v_L}{\sqrt{2}}\frac{h_2+h_3}{2}\,.
 \eeq
 One can see that the leading mass splitting between $E_+$ and $E_-$, apart from the phase convention, comes from the SM Higgs Yukawa interaction, $h_{2,3}$,  and to a very good approximation,
 \beq
\label{eq:UsymB}
V_{L/R}\simeq V^B \equiv \left(
    \begin{array}{ccc} 1&0&0\\
    0&\frac{1}{\sqrt{2}} & \frac{1}{\sqrt{2}}  \\
    0& -\frac{1}{\sqrt{2}}&\frac{1}{\sqrt{2}} \\
    \end{array}
  \right)\,.
\eeq

In the basis $\{\nu_L, N_{1L}, N_{2R}^c\}$, the neutrino mass matrix is
\beq
\label{eq:iss}
{\cal M}_N =\frac{v_L}{\sqrt{2}}  \times \left(  \begin{array}{ccc}
                       0 & y_1\epsilon_t & f_1^* \\
                       y_1\epsilon_t & 0 & f_3^*\\
                       f_1^*& f_3^*& y_2\epsilon_t \\
                     \end{array}
\right)
\eeq
and $\epsilon_t= \frac{v_t}{v_L}<\epsilon_v$. Again, we consider the case that $f_1\ll1$ and $y_1\sim y_2 =y$. The eigenvalues can be worked out to be around $ (y\epsilon_t/f)^3$, $-1+(y\epsilon_t/2f)$, and $1+(y\epsilon_t/2f)$ in units of $ f v_L/\sqrt{2}$. It is natural to identify the first term as the mass of the active neutrino. For $y v_t \sim 0.1$GeV and $f v_L\sim 3$TeV, the resulting active neutrino mass is about $(y v_t)^3/(f v_L)^2\sim 0.1$ eV. From electroweak precision measurements
we expect $v_t \lesssim {\cal O}(1)$ GeV. We see that the
desired neutrino mass can be obtained without much tuning of the Yukawa couplings.

 Notice that the neutrino mass matrix given in Eq.(\ref{eq:iss}) is of the inverse seesaw type\cite{ISS1,ISS2}, and a review can be found in \cite{ISSr}. The novel feature here is that we do not require ad hoc addition of the SM singlet leptons. The additional
leptons are dictated by anomaly cancelation and are SM doublets.

\section{3-generation lepton masses}
One can extend the above to the realistic 3-generation case.
Without losing any generality, we can start with the basis that the Yukawa couplings for  $\overline{N_{2R}}N_{1L}$ are diagonal.
And we can go to the basis where the SM charged leptons are in their mass eigenstates by bi-unitary transformation among the $e_R$ and $e_L$.
Similarly, we have the freedom to start with  diagonal $ (3\times 3)$ $\bar{E}_{1R} E_{2L}$ and $\bar{E}_{2R}E_{1L}$   sub-matrices.
\subsection{Charged lepton mass matrix}
For simplicity, let's consider that   $f_{1,2}=0$, $f_{3,4} \sim f$,  and the heavy charged lepton are roughly degenerate.
Then, in the basis $(\mathbf{e},\mathbf{E_1},\mathbf{E_2})$ where each entry is a 3-vector in family space, the most general $(9\times 9)$ mass matrix for charged leptons looks like
\beq
{\cal M}_C= \frac{v_L}{\sqrt{2}}\left(
  \begin{array}{ccc}
 h_1 \epsilon_v & \mathbf{0} & \mathbf{0}\\
  \mathbf{0} & h_2 \epsilon_v & f\cdot \mathbf{1}+\delta_1 \\
  \mathbf{0}& f \cdot \mathbf{1}+\delta_2 & h_3 \epsilon_v
  \end{array}
\right)
\eeq
where $h_1$ and $\delta_{1,2}$ are $3\times 3$ diagonal matrices and $\mathbf{1}$ is the unit matrix.
For convenience, $\delta_{1,2}$ which encodes the small splitting of the heavy charged leptons
are separated out from the leading term.
One can first perform a rotation among the heavy charged leptons by $U=\mathbf{V}_B$, which is a $(9\times 9)$ generalization of Eq.(\ref{eq:UsymB}).
Then the small perturbation can be separated from the leading order mass eigenvalues,
\beqa
U^T.{\cal M}_C.U &=& {\cal M}_C^{(0)}+ \Delta {\cal M}_C\nonr\\
{\cal M}^{(0)}_C &=&  \left( \begin{array}{ccc}
  \mbox{diag}(m_e,m_\mu,m_\tau) & \mathbf{0}&\mathbf{0}\\ \mathbf{0} & -\frac{v_L}{\sqrt{2}}\left[ f\cdot\mathbf{1} +\frac{1}{2}(\delta_1+\delta_2)\right] & \mathbf{0}\\  \mathbf{0}&\mathbf{0} & \frac{v_L}{\sqrt{2}}\left[ f\cdot\mathbf{1} +\frac{1}{2}(\delta_1+\delta_2)\right]   \end{array}\right)\,,\nonr\\
\Delta {\cal M}_C &=&  \frac{v_L}{2\sqrt{2}} \left(
  \begin{array}{ccc}
  \mathbf{0} & \mathbf{0} & \mathbf{0} \\
  \mathbf{0} & (h_2+h_3) &  (h_2-h_3+\delta_1-\delta_2)\\
  \mathbf{0} &  (h_2-h_3-\delta_1+\delta_2) &(h_2+h_3)
  \end{array}
\right)\,.
\eeqa
One can further diagonalize the diagonal $(3\times 3)$ block, $(h_2+h_3)$, by a bi-unitary transformation such that $ V_L^\dagger\cdot(h_2+h_3)\cdot V_R =h_{diag}$.
Or by using
\beq
U_R= U\cdot \mbox{diag}\{\mathbf{1},V_R,V_R\}\,,\,  %
U_L= U\cdot \mbox{diag}\{\mathbf{1},V_L,V_L\}\,,
\eeq
so that
\beqa
U_L^\dagger.{\cal M}_C.U_R &=& \mbox{diag}\{\mathbf{M}_e,\mathbf{M}_-,\mathbf{M}_+\}+ \Delta {\cal M}'_C\nonr\\
\mathbf{M}_e &=&\mbox{diag}(m_e,m_\mu,m_\tau)\,,\,\,
\mathbf{M}_\pm = \pm \frac{v_L}{\sqrt{2}}\left[ f\cdot\mathbf{1} +\frac{1}{2}(\delta_1+\delta_2\pm h_{diag})\right]\,,\nonr\\
\Delta {\cal M}'_C &=&  \frac{v_L}{2\sqrt{2}} \left(
  \begin{array}{ccc}
  \mathbf{0} & \mathbf{0} & \mathbf{0} \\
  \mathbf{0} & \mathbf{0} &  V_L^\dagger\cdot(h_2-h_3+\delta_1-\delta_2)\cdot V_R\\
  \mathbf{0} &  V_L^\dagger\cdot(h_2-h_3-\delta_1+\delta_2)\cdot V_R & \mathbf{0}
  \end{array}
\right)\,.
\eeqa
It is clear that the 6 heavy charged leptons will form 3 nearly degenerate pairs.  And as in the 1-generation case, the
mass splitting for each pair is mainly controlled by $h_{2,3}$.
Moreover, they decouple from the SM charged leptons.
\subsection{Neutral lepton mass matrix}
Using the notation of the charged leptons and factor out the common mass,  we write the general  $(9\times 9)$ neutrino mass matrix as
\beq
{\cal M}_N= \frac{f v_L}{\sqrt 2} \widetilde{{\cal M}}_N\,,\,\,  \widetilde{{\cal M}}_N=\left(
  \begin{array}{ccc}
  \mathbf{0} & \epsilon_1 & \mathbf{0}\\
  \epsilon_1^T & \mathbf{0} & \mathbf{1}+\delta \\
 \mathbf{ 0} & \mathbf{1}+\delta & \epsilon_2
  \end{array}
\right)
\eeq
where $\epsilon_2$ is a symmetric $3\times 3$ matrix  with elements $\{\epsilon_2\}_{ij}=\{\epsilon_2\}_{ji}\sim {\cal O}(y_2 v_t/v_L)$ , $\epsilon_1$ is a general $3\times 3$ matrix  with elements $\{\epsilon_1\}_{ij} \sim {\cal O}(y_1 v_t/v_L)$, and now $\delta$ is a $3\times 3$ diagonal matrix $\delta=\mbox{diag}(0,\delta_1,\delta_2)$, $\delta_{1,2}\ll 1$, to accommodate the small non-degeneracy among the three heavy $N$s.
First, the leading mass diagonalization can be made by the same rotation $\mathbf{V}_B$, similar to Eq.(\ref{eq:UsymB}), as in the charged lepton case.
This results in a symmetric $3\times 3$ matrix,
 $\delta_3\equiv -\delta +\epsilon_2/2$, in the diagonal blocks as the perturbation.
Assume there exists an orthogonal $3\times 3$ transformation $V$, such that  $V^T\cdot \delta_3\cdot V=\mbox{diag}\{a_1,a_2,a_3\}\equiv \delta_4$, and $|a_{1,2,3}|\sim {\cal O}\left(\sqrt{\delta^2+\epsilon_2^2} \right)\ll 1$.
Then by using $U=\mathbf{V}_B\cdot \mbox{diag}\{\mathbf{1},V,V\}$, the re-scaled neutrino mass matrix can be brought into the following form
\beqa
(U)^T.\widetilde{{\cal M}}_N.U &=& {\cal M}^{(0)}+ \Delta {\cal M}\nonr\\
{\cal M}^{(0)} &=&  \left( \begin{array}{ccc}
 \mathbf{ 0} & \mathbf{0}&\mathbf{0}\\ \mathbf{0} & -\mathbf{1}+\delta_4 & \mathbf{0}\\  \mathbf{0}&\mathbf{0} & \mathbf{1}+\delta_4   \end{array}\right)\,,\,\,
\Delta {\cal M} = \left(
  \begin{array}{ccc}
  \mathbf{0} & y & y \\
  y^T & \mathbf{0}  &  - z\\
   y^T  & -z &  w
  \end{array}
\right)
\eeqa
where $y= \frac{\epsilon_1 \cdot V}{\sqrt{2}}\sim {\cal O}(\epsilon_1)$, $z= \frac{V^T \cdot\epsilon_2\cdot V}{2}\sim {\cal O}(\epsilon_2)$,
and $w=2 V^T\cdot \delta\cdot V\sim {\cal O}(\delta)$.
One can see that after this rotation, the leading mass eigenstates are nothing but the Cartesian basis.
By the standard perturbation techniques, it is easy to see that the SM neutrinos will acquire nonzero masses at the second order perturbation.
For example, at this order,
\beqa
m_1 \left(\frac{f v_L}{\sqrt 2}\right)^{-1} &=& \sum_{i=2}^9 { (\Delta {\cal M}_{1i})^2 \over 0- {\cal M}^{(0)}_{ii} }
= - \sum_{j=1}^3 \left[ {(y_{1j})^2\over -1+a_j} + {(y_{1j})^2\over 1+a_j} \right]
\simeq 2 \sum_{i=1}^3(y_{1i})^2 a_i\,,
\eeqa
and it is indeed of the order of ${\cal O}(\epsilon_1^2 \epsilon_2)$ as in the 1-generation case.

\begin{figure}[h!]
 \centering
 \includegraphics[width=0.65\textwidth]{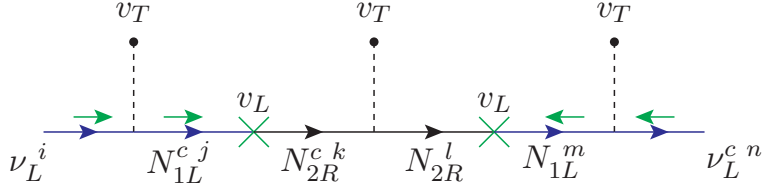}
\caption{Diagrammatic representation of the $\epsilon^3$-suppression for the active neutrino masses.
Superscripts denote family indices. Upper(green) arrows denote flow of lepton charge.}
\label{fig:numass}
 \end{figure}

 The active neutrino masses can also be understood diagrammatically. By integrating out the heavy $N$, the corresponding Feynman diagram in the weak basis, displayed in Fig.\ref{fig:numass}, and can be seen to give the same conclusion.
It also reveals that the low energy effective operator for active neutrino mass is not given by the Weinberg operator.
 If we assume a hierarchy that $v_L \gg v \gg v_t$, and  $T$ is the only beyond SM degree of freedom left below $v_L$, the active neutrino masses are generated  by a dimension six operator $O_6= \frac{c}{(\Lambda_L)^2}(\ovl{l^c_L} T^\dagger l_L)Tr(T^\dagger T)$ where $c$ is a constant and $\Lambda_L$ is the lepton number breaking scale related to $v_L$. After $T$ picks up a VEV, $v_t$, the neutrino mass is given by $m_\nu \simeq \frac{c v_t^3}{\Lambda_L^2}$. It is also clear that
 $O_6$ has a higher dimension than the Weinberg operator. Together with the fact that $v_t\ll v$., they allow the lepton breaking scale to be much lower than the usual type I seesaw mechanism.

 Now, the  upper-left $(3\times 3)$ sub-matrix, denoted as $\mathbf{N}_\nu$, of $U$ for active neutrinos is in general non-unitary,
 $\mathbf{N}_\nu \mathbf{N}_\nu^\dagger \neq 1$. This non-unitarity will result in various observable effects.
 However, one expects  that the off-diagonal elements of $|\mathbf{N}_\nu \mathbf{N}_\nu^\dagger|$ are of the order of ${\cal O}(\epsilon_1^2) \sim 10^{-6} \times (v_t/\mbox{GeV})^2 \times (\mbox{TeV}/v_L)^2$, which is roughly below the current experimental limits, $\lesssim 10^{-5}$\cite{nonU_1,nonU_2}. Therefore, we will leave the comprehensive study of these precision tests to future work.

\section{Neutrino oscillations and data fitting}
First, we provide a simple, realistic solution which can accommodate the neutrino data. Then we move on to the more general numerical survey where the solutions will be fed into the later study of lepton flavor changing processes.

To simplify the discussion,  we assume that the heavy $N$'s are degenerate($\delta=0$), $y_2\propto \mathbf{1}$, and all the Yukawa couplings in $y_1$ are of the same order and there is no hierarchy among them.
The $(9\times 9)$ mass matrix looks like
\beq
{\cal M}_N =  v_L \left(  \begin{array}{ccc}
                       \mathbf{0} & \epsilon y_1 & \mathbf{0} \\
                       \epsilon y_1^T& \mathbf{0} & \mathbf{1}\\
                       \mathbf{0}& \mathbf{1}& \epsilon y_2 \\
                     \end{array} \right)\,,
\eeq
 where $\epsilon\sim {\cal O}(v_t/v_L)$ is an unknown overall constant which controls the amplitude of perturbation and the elements of $y_1$ are of $\sim {\cal O}(1)$.
As discussed previously, in the leading approximation, the $(3\times 3)$ active neutrino mass matrix reads
\beq
{\cal M}^\nu_{ij} \sim \epsilon^3 v_L \{y_1\}_{i\alpha} \{y_2\}_{\alpha\beta} \{y_1\}_{j \beta}
\sim y_2 \epsilon^3 v_L \{y_1\}_{i\alpha} \{y_1\}_{j \alpha}\,.
\eeq
If $y_1$ is highly democratic, namely,
\beq
y_1 \propto \mathbf{I}_c \equiv  \left(  \begin{array}{ccc} 1&1&1\\ 1&1&1\\1&1&1\\ \end{array} \right)\,,
\eeq
the resulting active neutrino mass matrix also has the pattern ${\cal M}^\nu \propto \mathbf{I}_c$
which is of rank one and it has two zero eigenvalues. It naturally leads to the normal hierarchical neutrino masses.
Taking into account the data, the realistic mass matrix for normal hierarchy(NH) instead takes the form
\beq
{\cal M}^\nu \sim \left(  \begin{array}{ccc} 0.1&0.1&0.1\\ 0.1&1&1\\0.1&1&1\\ \end{array} \right) \times (0.03) \mbox{ eV}
\eeq
if $m_1\simeq 0$, and, to simplify the discussion we set $\delta_{CP}=0$.
A simple solution to arrive such pattern is
\beq
y_1 \sim \frac{y}{\sqrt{3}} \left(  \begin{array}{ccc} -0.3&0.3&0.3\\ 1&1&1\\1&1&1\\ \end{array} \right)\,,\,\,
y_2\sim y \mathbf{1}
\eeq
which has the apparent $\mu-\tau$ symmetry.
This can be realized in the extra-dimensional models by arranging the amount of overlap in higher dimensional fermion wavefunctions,  see for example \cite{RS1_1,RS1_2,SF}.

On the other hand, a more subtle construction of $y_1$ is required to accommodate the inverted hierarchy( IH ) case.
For example, if $m_3\simeq 0, \delta_{CP}=0$, the following realistic neutrino masses matrix
\beq
{\cal M}^\nu \sim \left(  \begin{array}{ccc} 1.6&-0.2&-0.2\\ -0.2&0.9&-0.8\\-0.2&-0.8&0.8\ \end{array} \right) \times (0.03) \mbox{ eV}
\eeq
can be generated by
\beq
y_1 \sim y \left(  \begin{array}{ccc} -0.7&-0.7&-0.7\\ 0.3&0.6&-0.7\\-0.1&-0.4&0.8\\ \end{array} \right)\,,\,\,
y_2\sim y \mathbf{1}\,.
\eeq

For both NH and IH cases, $ y^3 v_t^3 /v_L^2 \sim 0.03 {eV}$. Taking $v_t =1$GeV and $v_L=1(5)$TeV, we have $y \sim 0.03 (0.09)$.
This simple solution with $y_2\sim y \cdot\mathbf{1}$ gives us a rough idea of the Yukawa coupling strengths.

For the realistic data fitting, we perform a comprehensive numerical scan
with the working assumption that $|(y_2)_{ij}|\simeq y_2$ and that the heavy $N$'s are nearly degenerate.
These assumption can be relaxed giving rise to more free parameters to fit the data.
Moreover, the Yukawa couplings are taken to be complex in the numerical study to accommodate the nonzero CP phase, $\delta_{CP}$ which current data give a hint of.
However, it is clear that the resulting neutrino mass is about $m_\nu\sim (y_1^2 y_2 v_t^3/M_N^2)$.
We adopt the following $3\sigma$ ranges from \cite{NuFit} for the neutrino oscillation parameters.
For NH,
\beqa
&&31.42^\circ<\theta_{12}<36.05^\circ\,,\,40.3^\circ<\theta_{23}<51.5^\circ\,,\,8.09^\circ<\theta_{13}<8.98^\circ\,,\,
144^\circ<\delta_{CP}<374^\circ\,,\nonr\\
&&\Delta m_{21}^2=(6.8-8.02)\times 10^{-5}\mbox{ eV}^2\,,
\Delta m_{31}^2=(2.399-2.593)\times 10^{-3}\mbox{ eV}^2\,.
\eeqa
As for the IH case, the corresponding $3\sigma$ ranges are:
\beqa
&&31.43^\circ<\theta_{12}<36.06^\circ\,,\,41.3^\circ<\theta_{23}<51.7^\circ\,,\,8.14^\circ<\theta_{13}<9.01^\circ\,,\,
192^\circ<\delta_{CP}<354^\circ\,,\nonr\\
&&\Delta m_{21}^2=(6.8-8.02)\times 10^{-5}\mbox{ eV}^2\,,
\Delta m_{31}^2=-(2.369-2.562)\times 10^{-3}\mbox{ eV}^2\,.
\eeqa

The lightest neutrino masses, $m_{\mbox{\tiny lightest}}$, for both NH and IH are allowed to vary in the range between $10^{-4}$eV and $0.2$eV so that the cosmological bound,  $\sum_j m_j<0.57$eV at $95\%$ C.L.(from CMB spectrum and gravitational lensing data only)\cite{numassCos}\footnote{This upper limit
has been improved to $\sum_j m_j<0.44$eV ( or $m_{\mbox{\tiny lightest}}\lesssim 0.15$eV ) at $95\%$ C.L.\cite{Planck_2018} recently. },  can be met.
Once $m_{\mbox{\tiny lightest}}$ is fixed, $m_{1,2,3}$ can be determined from the measured mass squared differences.
Then the effective active neutrino mass matrix can be obtained by
\beq
{\cal M}_\nu = U_{PMNS}^* \cdot\mbox{diag}(m_1,m_2,m_3)\cdot U_{PMNS}^\dagger\,.
\eeq
In the standard parametrization, the rotation matrix is given by\footnote{In general $U_{PMNS}=U^\dagger_{c.l.}U$ where $U_{c.l.}$ is the unitary matrix that diagonalizes the
SM charged leptons mass matrix and $U$ is the neutrino rotation matrix. In the limit that
the $\mathbf{E}$ decouples from the $\mathbf{e}$ we can take $U_{c.l.}=\mathbf{1}$.}
\beq
\label{eq:definition_of_pmns}
 U_{\mathrm{PMNS}}=\left(\begin{array}{ccc} 1 & 0 & 0\\
0 & c_{23} & s_{23}\\
0 & -s_{23} & c_{23}\end{array}\right)\left(\begin{array}{ccc}
c_{13} & 0 & s_{13} e^{-i\delta} \\
0 & 1 & 0 \\
-s_{13} e^{i\delta} & 0 & c_{13}\end{array}
\right)\left(\begin{array}{ccc}
c_{12} & s_{12} & 0 \\
-s_{12} & c_{12} & 0 \\
0 & 0 & 1\end{array}\right)
\begin{pmatrix}
1 & 0 & 0
\\
0 & e^{  i \alpha_{21}/2} & 0
\\
0 & 0 & e^{  i \alpha_{31}/2}
\end{pmatrix}
\eeq
where the shorthand $s_{12}\equiv \sin\theta_{12}$ and the like are used.
Each element of the $y_2$ Yukawa matrix is a random number in between $0.7$ and $1.3$ times an overall unknown factor $y_2$ with either sign.
And we require that the ratio of the largest to the smallest absolute values in $y_1$ to be smaller than $10$.
About $10^5$ such solutions are prepared for both NH and IH cases.
The realistic Yukawa coupling configurations can be used for predicting the lepton flavor violating processes.
The results will be displayed in the next section.

\section{  $\mu\ra e \gamma$, and $a_\mu$}
\begin{figure}[h!]
 \centering
 \includegraphics[width=0.5\textwidth]{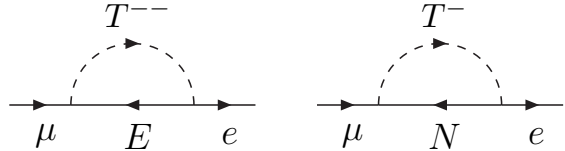}
\caption{Feynman diagrams for $\mu\ra e \gamma$ transition, the photon can attach to any charged line.}
\label{fig:MEG}
 \end{figure}

With this rich exotic leptons introduced for anomaly cancelations it is important to examine
the constraints from charged lepton flavor violation (CLFV) searches.
The $\mu\ra e \gamma$ process can be mediated by triplet running in the loop, see Fig.\ref{fig:MEG}.
In the lepton mass basis, with the $\mathbf{V}_B$ rotation, the triplet coupling can be approximated as
\beqa
\left. \frac{(y_1)_{ij}}{\sqrt{2}} \right\{ && \bar{T}_0 \bar{\nu^c}_i(N_{j+} +N_{j-}) + T_{++} \bar{e}^c_i(E_{j+}+E_{j-} )  \nonr\\
 && \left. +\frac{1}{\sqrt{2}} T_+ \left[ \bar{\nu^c}_i(E_{j+}+E_{j-} ) +\bar{e}^c_i (N_{j+} +N_{j-}) \right]
\right\} +h.c.
\eeqa
where $i,j=1,2,3$ are the generation indices, and $\pm$ denote the different mass eigenstates within each generation.

\begin{figure}[h!]
 \centering
 \includegraphics[width=0.4\textwidth]{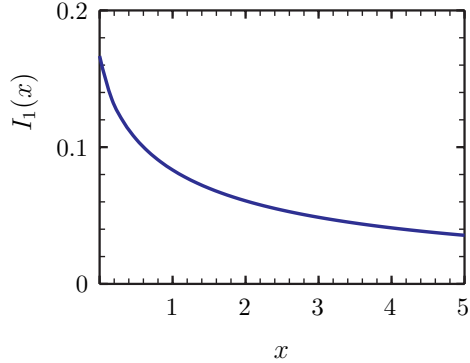}
\caption{The loop function $I_1(x)$.}
\label{fig:F1}
 \end{figure}

The 1-loop contributions  can be calculated to be
\beqa
\Delta a_\mu = -\sum_{i=1}^6 {|(y_1)_{\mu i}|^2 m_\mu^2 \over 32\pi^2}  \left[ {I_1\left((\tau_i^{--})^{-1}\right) \over  M_{E_i}^2}
+{2 I_1\left(\tau_i^{--}\right)  \over  M_{T_{--}}^2}
+ \frac{1}{2}{I_1\left(\tau_i^N\right) \over  M_{T_-}^2}\right]
\label{eq:mu_g-2}
\eeqa
where $\tau_i^{--}\equiv \frac{M_{E_i}^2}{M_{T_{--}}^2}$, $\tau_i^N \equiv \frac{M_{N_i}^2}{M_{T_{-}}^2} $, and
\beq
I_1(x)=\int^1_0 dz {z(1-z)^2 \over 1-z+ x z}= {1\over 6(1-x)^4}\left[1-6x+3x^2+2x^3-6x^2 \ln x\right]\,.
\eeq
See Fig.\ref{fig:F1} for the plot of this function.
When $x\ll 1$ and $x\sim 1$, the loop function can be expanded as
\beqa
I_1(x)&\simeq& \frac{1}{6}-\frac{x}{3}-\left(\frac{11}{6}+\ln x\right)x^2+{\cal O}(x^3)\,,\\
&\simeq& \frac{1}{12}-\frac{x-1}{20}+\frac{(x-1)^2}{30}+{\cal O}((x-1)^3)\,.
\eeqa
The first term in the square bracket of Eq.(\ref{eq:mu_g-2}) is the contribution where the photon attaches to the heavy charged lepton. The second and third terms are the
contributions where the photon attaches to the $T_{--}$ and $T_-$, respectively.
Because of the electric charge, the $T_{--}$ contribution has an extra factor 2.
Note also the one half factor associated with the $T_-$ contribution which is due to the extra $1/\sqrt{2}$ factor in the singly charged triplet-fermion vertex coupling.
Moreover, assuming that $M_E\sim M_N \sim M$ ( so that all $I_1\sim 1/12$) the $(g-2)_\mu$ can be related to the neutrino mass
 and eliminating the $y_1$ dependence,
\beq
\Delta a_\mu  \sim -{3  m_\mu^2 m_\nu \over 16 \pi^2 y_2 v_t^3}\left[ I_1\left((\tau_i^{--})^{-1}\right)
+2 I_1\left(\tau_i^{--}\right)\frac{M^2}{M_T^2}
+\frac{1}{2} I_1\left(\tau_i^N\right)\frac{M^2}{M_T^2}  \right]
\sim -10^{-15}\frac{(\mbox{GeV})^3}{y_2 v_t^3}
\eeq
which is negligibly small.

Similar calculation can be carried out for the $\mu\ra e \gamma$ dipole transition amplitude.
\beqa
 {\cal M}^\mu_{\mu\ra e\gamma}
&=& A_{\mu e}\, \overline{e}(p_2) \left(-i\sigma^{\mu q}\right) \hat{R} \mu(p_1)\,,\nonr\\
A_{\mu e}&=& \sum_{i=1}^6 { e (y_1)_{\mu i}(y_1)_{e i}^* m_\mu \over 64\pi^2}\,\, \left[ {I_1\left((\tau_i^{--})^{-1}\right) \over  M_{E_i}^2}
+{2 I_1\left(\tau_i^{--}\right)  \over  M_{T_{--}}^2}
+\frac{1}{2} {I_1\left(\tau_i^N\right) \over  M_{T_-}^2}\right]\,,
\eeqa
where $q^\mu\equiv (p_2-p_1)^\mu$ is the photon 4-momentum, and $\hat{R}=(1+\gamma^5)/2$.
Then, the branching ratio is\cite{Chang:2005ag}
\beq
Br(\mu\ra e\gamma )= {12 \pi^2  A_{\mu e}^2 \over G_F^2 m_\mu^2}\sim {27 \alpha\over 64\pi} {(y_1)^4 (3.5 I_1)^2 \over G_F^2 M^4}
\sim 10^{-13} \left({\mbox{TeV} \over M}\right)^4 \left({(y_1) \over 0.01}\right)^4\,.
\eeq
Or, assuming that $M_E\sim M_N \sim M$, the LFV process can be related to the neutrino mass,
\beq
A_{\mu e} \sim {3 e m_\mu m_\nu \over 32 \pi^2 y_2 v_t^3}\left[ I_1\left((\tau_i^{--})^{-1}\right)
+2 I_1\left(\tau_i^{--}\right)\frac{M^2}{M_T^2}
+\frac{1}{2} I_1\left(\tau_i^N\right)\frac{M^2}{M_T^2}  \right]\,.
\eeq
Note that the mass squared of heavy leptons in numerator and denominator cancel out,
and the branching ratios is not very sensitive to the masses of heavy degree of freedom.

\begin{figure}[htb]
 \centering
 \includegraphics[width=0.7\textwidth]{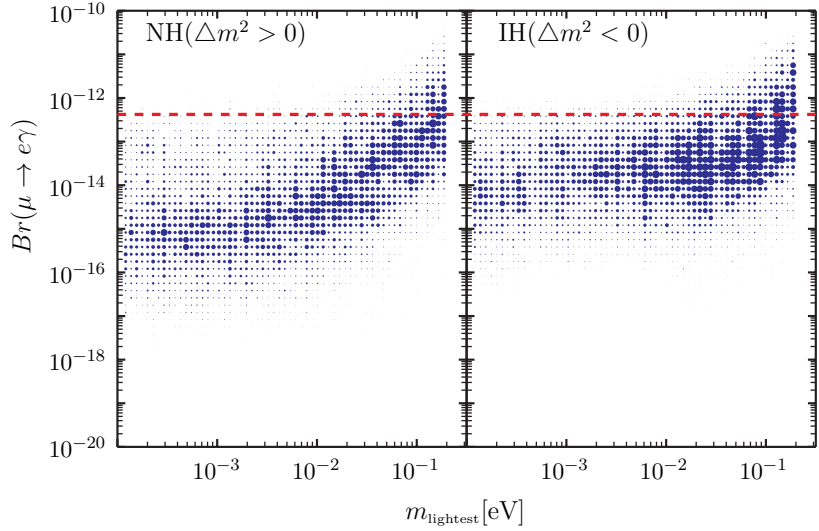}
\caption{The log-log plot of $Br(\mu\ra e\gamma)$ v.s. $m_{lightest}(eV)$ in our model.
NH/IH is on the left/right panel. The red dash line indicates the current experimental bound\cite{MEG}. The radius of each dot is proportional to the number of found solutions in the corresponding $Br-m_{\mbox{\tiny lightest}}$ cell.
 We take $y_2 v_t^3=1(GeV)^3$ and $M_N=M_E=M_T$.}
\label{fig:LFVMeA}
 \end{figure}

Comparing to the most recent bound, $Br(\mu\ra e \gamma)<4.2\times 10^{-13}$ at 90\% C.L.\cite{MEG},
our  numerical results are shown in Fig.\ref{fig:LFVMeA}.
As can be seen from the plot, it is easier to find the solutions for larger $m_{\mbox{\tiny lightest}}$ in the IH case.
For $y_2 v_t^3=1(GeV)^3$,  the $\mu\ra e\gamma$ branching ratio is right below the current experimental limit for $m_{\mbox{\tiny lightest}}\lesssim 10^{-2}$eV.
Note that the branching ratios have  lower bounds, around $\sim 10^{-16}/10^{-15}$ for NH/IH case with $y_2 v_t^3=1(GeV)^3$.
Therefore, for this model to admit a realistic solution which accommodates simultaneously the neutrino oscillation data and the current $\mu\ra e\gamma$ bound, the predicted lower bounds must stay below the experimental limit.  It is required that
\beq
{ 1(\mbox{GeV})^3\over y_2 v_t^3 } < 64.807(20.493)
\eeq
for NH(IH).
Thus, we arrive an interesting lower bound on the triplet VEV that
\beq
v_t > { 0.249(0.365)\over (y_2)^{1/3} }\,\, \mbox{GeV} >0.107(0.157) \mbox{GeV}
\eeq
for the NH(IH) case, where the ultima bound is obtained by taking the strong coupling limit $y_2=4\pi$.

From Eq.(\ref{eq:TL_triplet_mass}), this lower bound implies that the triplet mass is roughly below $\lesssim 8$TeV if $\kappa\sim v$.

\begin{figure}[h!]
 \centering
 \includegraphics[width=0.7\textwidth]{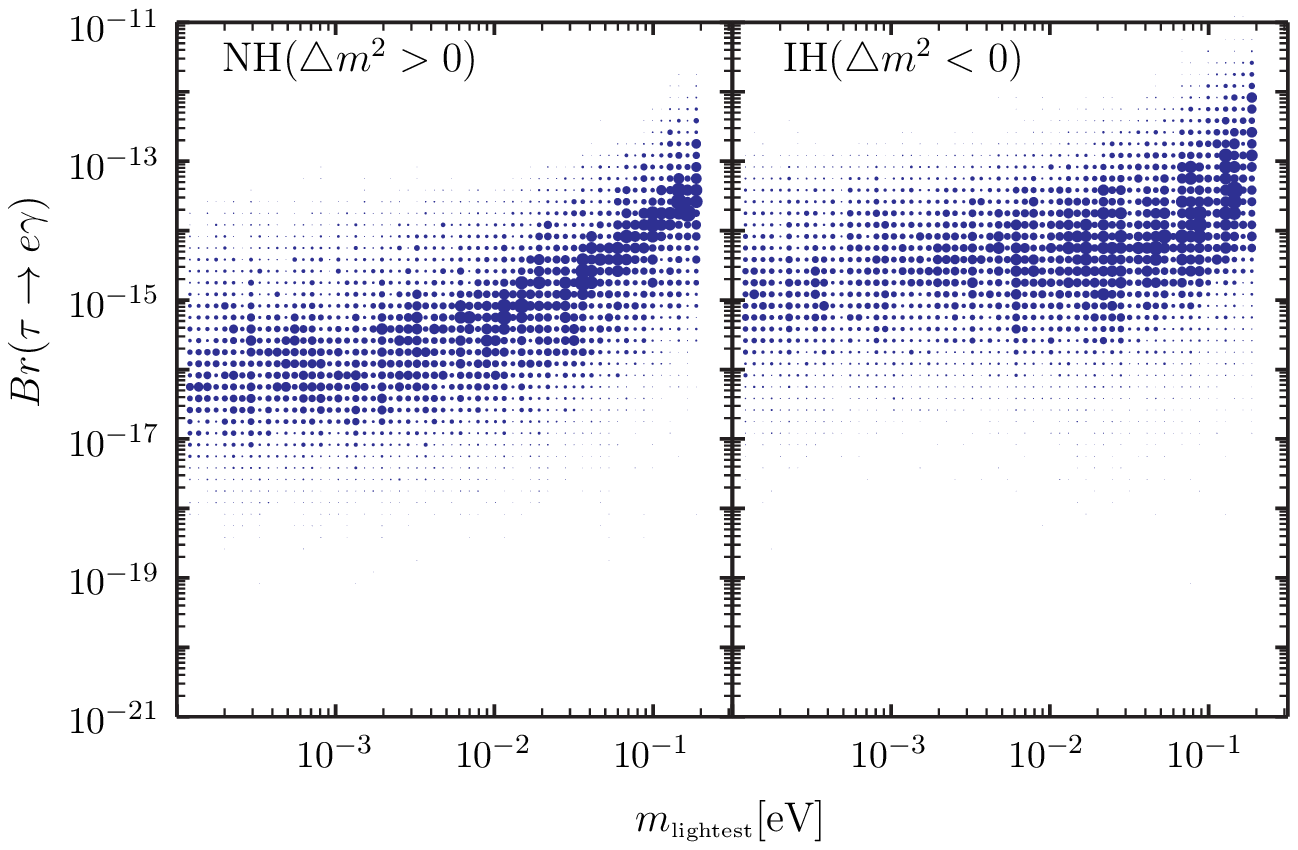}
 \\
  \includegraphics[width=0.7\textwidth]{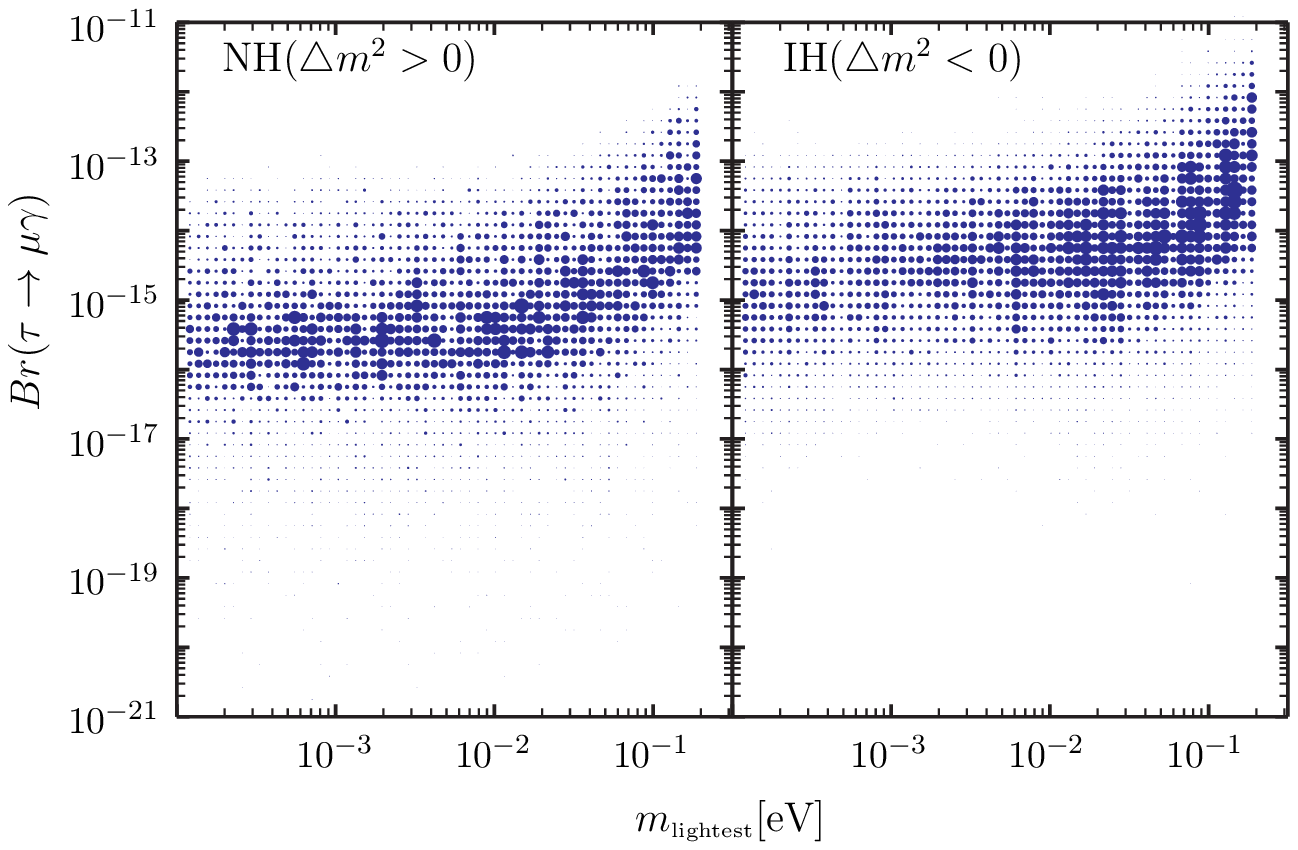}
\caption{ $Br(\tau\ra e\gamma)$ and $Br(\tau\ra \mu\gamma)$ v.s. $m_{\mbox{\tiny lightest}}$ in our model.
 We take $y_2 v_t^3=1(GeV)^3$ and $M_N=M_E=M_T$.}
\label{fig:LFVTlA}
 \end{figure}

Since in our model the triplet does not carry lepton number, there is no tree-level contribution to $\mu\ra 3 e$ and the similar $\tau$ decays.
The dipole induced $Br(\mu\ra 3e)$ will be small comparing to $\mu\ra e\gamma$. The ratio\cite{Chang:2005ag} is given by
\beq
{ Br(\mu\ra 3e ) \over Br(\mu\ra e\gamma )} ={2\alpha\over 3\pi}\left(\ln\frac{m_\mu}{m_e}-\frac{11}{8}\right)\simeq 0.7\times 10^{-2}
\eeq
which makes $Br(\mu\ra 3e)<3\times 10^{-15}$ in this model.
Similarly, the branching ratios of $\tau\ra l\gamma\, (l=e,\mu)$  are
\beq
Br(\tau\ra l\gamma) \simeq  {12 \pi^2  A_{\tau l}^2 \over G_F^2 m_\tau^2}\times Br(\tau\ra e \bar{\nu}_e \nu_\tau)
\eeq
and we adopt the measured  $Br(\tau\ra e \bar{\nu}_e \nu_\tau)=17.82\%$\cite{PDG2016}.
The predicted branching ratios of $\tau\ra l \gamma$ in our model are displayed in Fig.\ref{fig:LFVTlA} which are much smaller than the current
experimental bound; $Br(\tau\ra e \gamma)< 3.3 \times 10^{-8}$ and $Br(\tau\ra \mu \gamma)< 4.4 \times 10^{-8}$ at 90\%C.L.\cite{PDG2016}.
Note that in the IH cases the two have same statistics which is  due to the complex conjugated pair solutions to the $y_1$ Yukawa for a given $m_{\mbox{\tiny lightest}}$ and $U_{PMNS}$.
As pointed out in \cite{CZB}, the double ratios, for example, $Br(\mu\ra e\gamma)/Br(\tau\ra e\gamma)$, are independent of the unknown parameters
$y_2, v_t$ and the masses of the heavy degrees of freedom. They are complementary handles to the long baseline experiments for determining the type of neutrino mass hierarchy. Unfortunately, we have not found any notable statistical  difference between the double ratios of NH and IH in this model.

\section{Triplets at colliders}
The phenomenology of the $Z_\ell$ and the
charged heavy leptons are the same as in (I), and we shall not repeat them here. The triplets
are the new players and we will discuss their signatures at the LHC below. We start with a list of their dominant decay modes.
\subsection{Decays of the triplet}
Due to the gauge couplings and SSB, the triplet scalar can decay into (a) two SM gauge bosons collectively called $V$ (b) a lighter triplet partner plus a V, e.g. $T_{--}\ra T_-W^+$, and (c)  two light triplets, e.g. $T_{--}\ra 2T_-$. The later two require huge mass splitting or the rates are suppressed by $v_t$, thus can be ignored here\footnote{For example, $\Gamma(T\ra T_1 W)=G_F M^3 \lambda_{cm}^3(x_1,x_W)/(2\sqrt{2}\pi)$, where $x_1=(M_1/M)^2$ and $x_W=(M_W/M)^2$. However, there is no allowed phase space if the mass squared difference between $T$ and $T_1$ is at most $v^2$.}.
Therefore, $T\ra V_1V_2$ $ (V_{1,2}=W^\pm,Z)$  are the dominant decays since $T$ does not couple to  two SM fermions simultaneously in the weak basis.
This is very different from the cases of triplet with $l=2$ as discussed, for example, in \cite{HanT}.
Parameterizing the vertex $T V_1^\mu V_2^\nu$  Feynman rule as $i \kappa_{V_1,V_2} g^{\mu\nu}$, it's straightforward to calculate the
following decay widthes:
\beqa
\Gamma(T\ra V_1 V_2)&=& {|\kappa_{V_1,V_2}|^2 \over 16\pi M_T} \lambda_{cm}(x_1^2,x_2^2)\left[2 +\left({1-x_1^2-x_2^2\over 2x_1 x_2}\right)^2\right]\,,\\
\Gamma(T\ra V_1 V_1)&=& {|\kappa_{V_1,V_1}|^2 \over 32\pi M_T} \sqrt{ 1-4 x_1^2} \left[2 +\left({1-2 x_1^2\over 2x_1^2 }\right)^2\right]\,,\\
\Gamma(T\ra V_1 \gamma )&=& {3|\kappa_{V_1,\gamma}|^2 \over 16\pi M_T} \left( 1- x_1^2\right)\,,
\eeqa
where $x_i \equiv M_{V_i}/M_T$ and  $\lambda_{cm}(y,z)\equiv \sqrt{1+y^2+z^2-2y-2z-2yz}$.
The couplings are listed in Table. \ref{tab:TVV_vertex}.
\begin{table}[thb]
\begin{center}
\begin{tabular}{|c|c|c|c|c|c|}
\hline
Vertex & $\Re T_0 W^+W^-$ & $\Re T_0 Z Z$ & $T_{--}W^+W^+$ & $T_-W^+Z$ & $T_-W^+ P$\\ \hline
$\kappa_{VV}$ & $i g_2^2 v_t$  & $2i \frac{g_2^2}{c_W^2} v_t$  & $i \sqrt{2}g_2^2 v_t$  & $i \frac{g_2^2}{\sqrt{2}c_W}(1+s_W^2)v_t$ &
 $-i \frac{g_2 e}{\sqrt{2}} v_t $\\\hline
\end{tabular}
\caption{Feynman rules for $TVV$ vertices. The $g^{\mu\nu}$ factors are omitted. }
\label{tab:TVV_vertex}
\end{center}
\end{table}
The typical decay widthes for charged triplets are narrow, around ${\cal {O}}(10^{-2})$ MeV, for $v_t\sim 1$GeV and $M_T\sim 1$TeV. However, the charged triplet still decays promptly once produced. Moreover, the signal of triplet will be 4 fermion final state from the decay of two gauge bosons or 2 fermion plus a high energy photon.

On the other hand, if there is mixing between $\Re T_0$ and the Higgs boson, $ t_0$ can decay into fermion pairs. The two body decay width of $t_0$ is given by
\beq
\Gamma( t_0\ra f\bar{f})=\frac{|U_h^{12}|^2 G_F M_T}{ 4\pi \sqrt{2}} \sum_f  N_c m_f^2\left(1-\frac{4m_f^2}{M_T^2}\right)^{3/2}
\eeq
where $U_h$ is given in Eq.(\ref{eq:Uh}). This will be dominated by the $t\bar{t}$ final state if $M_T \gg M_t=174$ GeV.
LHC-1 gave a  bound on the SM signal strength that $\mu=1.09\pm 0.11$\cite{LHC_mu}, which implies that $|U_h^{12}|^2<0.13$ at $2\sigma$ level.
For $M_T=0.5(1.0)$TeV, the 2-body decay width has an upper bound $\Gamma( t_0\ra t\bar{t}) <8(36)$ MeV,
and $\Gamma( t_0\ra b\bar{b}) <0.57(1.1)$ MeV.
The mixing with the SM Higgs will also provide additional 2 gauge bosons decay widthes,
\beqa
\Gamma( t_0\ra W^+W^-)&=&\frac{|U_h^{12}|^2 G_F M_T^3} {32\pi \sqrt{2}}\sqrt{1-x_W}(4-4x_W+3x_W^2)\,,\nonr\\
\Gamma( t_0\ra ZZ)&=&\frac{|U_h^{12}|^2 G_F M_T^3}{ 64\pi \sqrt{2}}\sqrt{1-x_Z}(4-4x_Z+3x_Z^2)\,,
\eeqa
where $x_V\equiv 4M_V^2/M_T^2$. For $M_T=0.5(1.0)$TeV, $\Gamma( t_0\ra W^+W^-)\simeq 2\Gamma(t_0\ra ZZ)<5.3(42.6)$GeV.

Finally, we discuss the $t_0\ra 2h_{SM}$ decay.
Since $|U_h^{12}|\ll 1$, the relevant Lagrangian is roughly
\beq
\simeq \left[3\lambda_H v U_h^{12} +\frac{1}{2}(\lambda_1 v_t +\kappa) \right]h_{SM}^2 t_0
\eeq
and the $\kappa$ term dominates. We have
\beq
\Gamma( t_0\ra 2h_{SM})\simeq {\kappa^2 \over 32\pi M_T}\sqrt{1-x_H} = 0.172(0.096)\times\left({\kappa\over 100 \mbox{GeV}}\right)^2 \mbox{GeV}
\eeq
for $M_T=0.5(1.0)$TeV.
With non-zero mixing with Higgs, the two gauge bosons are still the dominant decay channel of $t_0$.
Moreover, for $M_{t_0}\gg M_Z$, the decay branching ratios of $Br(t_0\ra ZZ)\simeq 1/3$ and  $Br(t_0\ra W^+ W^-)\simeq 2/3$ are quite robust.

\subsection{Triplet production at hadron colliders}

As seen in the previous section, the production and decay of $t_0$ is very sensitive to its mixing with the SM Higgs.
We will start with the case that the mixing between $h,\Re T_0$ is negligible and focus on the production of the charged triplet at the collider.
The pair production at the LHC is mainly by the Drell-Yan processes through the $TTV$ vertices.
The gauge boson associated production cross section, $\sigma(pp\ra V T)$, is proportional to $v_t^2$ and negligible.
If ignoring the mixing and mass differences, $\sigma(pp\ra T_+T_{--})=\sigma(pp\ra T_0^* T_{-})$
and $\sigma(pp\ra T_-T_{++})=\sigma(pp\ra T_0 T_{+})$ for they have the same couplings and mediated by the s-channel $W$-exchange diagrams.
The cross sections at LHC14 for some typical triplet masses, listed in Table.\ref{tab:LHC_Tc_prod}, are evaluated by the program CalcHep\cite{CalcHep} with the CTEQ6l1\cite{CTEQ} PDF.

Note that  $pp\ra t\bar{t}W$ will be dominant SM background for $T_{--}T_{++}$.
 After applying proper cuts, a doubly charged of mass up to about $0.7$  TeV, and it decays mainly into di-boson,  can be probed at LHC14
 with an integrated luminosity of $300 fb^{-1}$\cite{HanT}. However, we defer a full study of the signal and proper treatment of the background to a future
 study.

\begin{table}[h!]
\begin{center}
\renewcommand{\arraystretch}{1.30}
\begin{tabular}{|c|c|c|c|c|}
\hline
 $M_T$ (TeV)& $T_{++}T_{--}$ &$T_{+}T_{-}$ & $T_+ T_{--}$ &$T_{++}T_-$  \\\hline
$0.5$ & $1.77$ & $0.179$ & $0.872$& $2.40$\\ \hline
$0.7$ & $0.345$ & $3.50\times 10^{-2}$&$0.157$& $0.496$\\ \hline
$1.0$ & $4.62\times 10^{-2}$ & $0.47\times 10^{-2}$&$1.93\times 10^{-2}$& $7.01\times 10^{-2}$\\ \hline
\end{tabular}
\caption{Charged Triplet bosons pair production cross sections(in $fb$) at the LHC14. Here we have neglected effects from $\Re H_0$ and $\Re T_0$ since it is small and give subleading contributions.}

\label{tab:LHC_Tc_prod}
\end{center}
\end{table}

\begin{figure}[h!]
 \centering
 \includegraphics[width=0.8\textwidth]{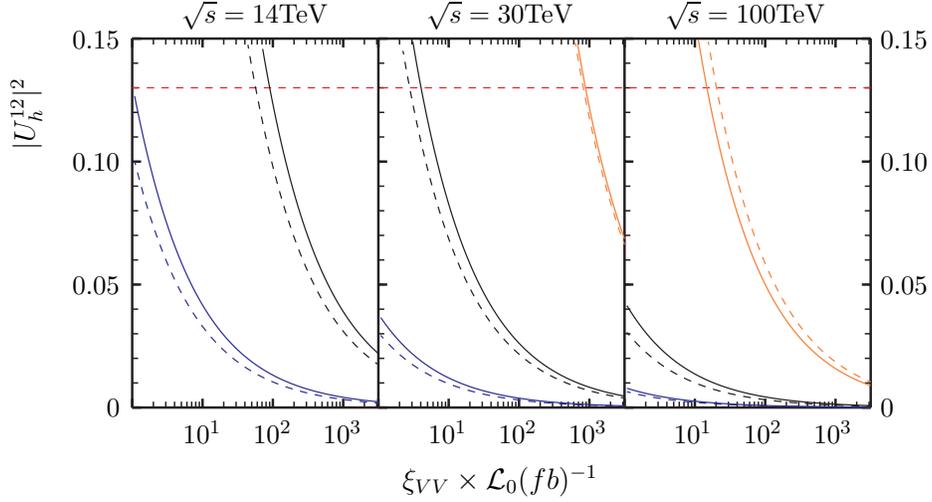}
 \caption{ The $5 \sigma$ limits of detecting a Triplet $t_0$ at the LHC. The curves are for $t_0$ decays into $W^+W^-$, and dashed ones are for $t_0\ra ZZ$.
 The color codes, (blue, black, orange), are for $M_T =(0.5,1.0,2.0)$ TeV, respectively.
 The horizontal dashed red line is the LHC-1 upper limit on $|U_h^{12}|^2$. }
\label{fig:LHC}
 \end{figure}

 In contrast, the real part of the neutral triplet\footnote{The single production of $A_0$ can be ignored for it has a  $v_t/v$ suppressing mixing with the Goldstone $G_0$, otherwise it couples only to one SM lepton doublet and one $L_1$, Eq.(\ref{eq:all_Yukawa}).}
 can be singly produced via gluon fusion through the mixing $(U_h^{12})$. Our estimates
of the production cross sections at the LHC and future hadron colliders are given in Table. \ref{tab:LHC_ggT0_prod}.
The SM backgrounds are estimated by evaluating the production cross section with the di-boson invariant mass in the $M_T\pm 50\mbox{GeV}$ range.
Derived from the numbers listed in Table. \ref{tab:LHC_ggT0_prod}, the $5\sigma$ limits  on the 2-dimensional $|U_h^{12}|^2$ and effective luminosity plane is shown in Fig.\ref{fig:LHC}.
The limit is determined by
\beq
{\mbox{Signal}\over \sqrt{\mbox{Background }}}=
{ \sqrt{{\cal L}_0\times \xi_{VV}} \times\sigma_S\times Br(t_0\ra VV) \over \sqrt{\sigma_{BG} } }=5
\eeq
where $\xi_{VV}$ is the efficiency of detection of $VV$ final states, and ${\cal L}_0$ is the integrated luminosity.
It can be seen that $t_0$ with a mass of $1$TeV and $|U_h^{12}|^2=0.05$ could be directly studied at the  LHC14 with $\sim 1 ab^{-1}$ effective luminosity.

\begin{table}[h!]
\begin{center}
\renewcommand{\arraystretch}{1.30}
\begin{tabular}{|c|c|c|c|c|}
\hline
$\sqrt{s}$(TeV)& $M_{t_0}$(TeV) & $\sigma(pp\ra t_0)$ &$\sigma(pp\ra W^+W^-)_{SM}$ & $\sigma(pp\ra ZZ)_{SM}$  \\\hline
$14$ &$0.5$ & $2.88\times 10^2$ & $2.56\times 10^3$ & $4.0\times10^{2}$ \\ \hline
$14$ &$1.0$ & $7.42$ & $1.49\times 10^2$ & $2.35\times10^{1}$ \\ \hline
$14$ &$2.0$ & $6.02\times 10^{-2}$ & $4.93$ & $0.879$ \\ \hline
$30$ &$0.5$ & $1.68\times 10^3$ & $7.27\times 10^3$ & $1.17\times10^{3}$ \\ \hline
$30$ &$1.0$ & $6.72\times 10^1$ & $5.49\times 10^2$ & $9.45\times10^{1}$ \\ \hline
$30$ &$2.0$ & $1.16$ & $3.56\times 10^1$ & $8.30$ \\ \hline
$100$ &$0.5$ & $1.63\times 10^4$ & $3.14\times 10^4$ & $5.27\times10^{3}$ \\ \hline
$100$ &$1.0$ & $9.73\times 10^2$ & $3.13\times 10^3$ & $4.26\times10^{2}$ \\ \hline
$100$ &$2.0$ & $3.03\times 10^1$ & $4.09\times 10^2$ & $1.40\times10^{2}$ \\ \hline
\end{tabular}
\caption{Gluon fusion neutral Triplet boson production cross sections(in $fb$) at the LHC and beyond.
Here we assume that $|U_h^{12}|^2=0.1$.
}
\label{tab:LHC_ggT0_prod}
\end{center}
\end{table}

\subsection{Triplet pair productions at the $e^+e^-$ machine}
\begin{figure}[h!]
 \centering
 \includegraphics[width=0.93\textwidth]{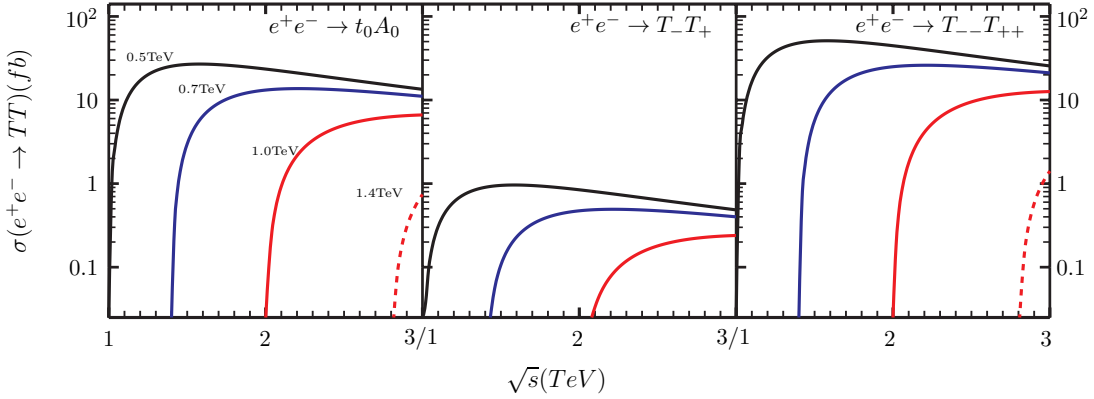}
  \caption{ The pair production cross sections ( in $fb$ ) for four different masses,$M_T=0.5,0.7,1.0,1.4$ TeV, v.s. $\sqrt{s}$ at the $e^+e^-$ collider. }
\label{fig:clic}
 \end{figure}

The triplet pair productions  at the $e^+e^-$ machine are mediated by the s-channel photon and Z diagrams.
 Ignoring the mixing between $\Re T_0$ and $h$,  the cross sections can be easily calculated to be:
\beqa
\sigma(e^+e^-\ra t_0 A)&=&\frac{\pi \alpha^2}{3 s} \frac{1}{s_W^2 c_W^2} \left( {1\over 1-\frac{M_Z^2}{s} }\right)^2\left(1-\frac{4M_T^2}{s}\right)^{3/2}\,,\\
\sigma(e^+e^-\ra T_{+} T_{-})&=&\frac{\pi \alpha^2}{3 s}\left(1- {\tan\theta_W\over 1-\frac{M_Z^2}{s} }\right)^2\left(1-\frac{4M_T^2}{s}\right)^{3/2}\,,\\
\sigma(e^+e^-\ra T_{++} T_{--})&=&\frac{4\pi \alpha^2}{3 s}\left(1+ {\cot 2\theta_W\over 1-\frac{M_Z^2}{s} }\right)^2\left(1-\frac{4M_T^2}{s}\right)^{3/2}\,.
\eeqa
The cross sections are displayed in Fig\ref{fig:clic}.
Note that the interference between photon and $Z$ contributions is destructive/constructive for $T_+T_-$/$T_{--}T_{++}$ production cross section.
Because the electric charge squared, $T_{\pm\pm}$ has the largest production cross section.
We use CalcHEP to estimate the SM backgrounds and find that they are about three orders of magnitude smaller than the signals, and thus negligible.

\section{EW precision, $\Delta S$ and $\Delta T$ from the exotic fermions and scalars}
\subsection{Tree-level $\rho-$parameter}

Since $T$ gets a VEV, $v_t$, the tree-level $\rho-$parameter is less than unit:
\beq
1+\alpha\Delta T_{tree}= \rho_{tree}= {v^2+2v_t^2 \over v^2+4v_t^2 } =1 - {2 v_t^2 \over v^2+4v_t^2 }\,.
\eeq
Therefore, the loop-induced $\Delta T_{loop}(>0)$ can be compensated by $\Delta T_{tree}(<0)$.
For $\Delta T= 0.08\pm0.12$\cite{PDG2016}, the $2\sigma$ range is
\beq
-0.16<\Delta T=\Delta T_{tree}+\Delta T_{loop} <0.32\,.
\eeq
The above only uses tree level contributions from the SM triplet  implies that $v_t<5.94$ GeV.
Combining with neutrino mass generation and the $\mu\ra e\gamma$ limit, we obtain the following interesting limit
\beq
 0.107<v_t< 5.94\; \mbox{GeV}\,.
 \label{eq:vtbound_no_loop}
\eeq

\subsection{Loop corrections}
Since anomaly cancelation mandates the addition of extra leptons, it is important to
know how quantum corrections to $\Delta T$ and $\Delta S$ from these new states will alter the above bound on $v_T$.

For each generation, the contributions from exotic leptons are \cite{Chang:2018vdd}
\beqa
\Delta T_{F_i} &=& {1\over 16\pi s_W^2 M_W^2} \left( M_{N_i}^2 +M_{E_i}^2 -2 {M_{N_i}^2 M_{E_i}^2 \over M_{N_i}^2 -M_{E_i}^2 } \ln\frac{M_{N_i}^2}{M_{E_i}^2} \right)\,,\\
\Delta S_{F_i} &=& \frac{1}{6\pi} \left(1 +\ln\frac{M_{N_i}^2}{M_{E_i}^2} \right)\,,
\eeqa
where $i=1,2$. From the triplet $T=(T_0,T_{-},T_{--})^T$, they are
\beqa
\Delta T_T &=& {1\over 8\pi s_W^2 M_W^2} \left( M_{T_0}^2 +M_{T_{-}}^2 -2 {M_{T_0}^2 M_{T-{-}}^2\over M_{T_0}^2 -M_{T_{-}}^2 } \ln\frac{M_{T_0}^2}{M_{T_{-}}^2} \right.\nonr\\
&& \left.M_{T_{-}}^2 +M_{T_{--}}^2 -2 {M_{T_{-}}^2 M_{T_{--}}^2\over M_{T_{-}}^2 -M_{T_{--}}^2 } \ln\frac{M_{T_{-}}^2}{M_{T_{--}}^2} \right)\,,\\
\Delta S_T &=& \frac{1}{3\pi} \ln\frac{M_{T_0}^2}{M_{T_{--}}^2}\,.
\eeqa
To simplify the discussion, we assume that all the exotic charged(neutral) leptons have the same mass $M_E(M_N)$, $T_-$ and $T_{--}$ are degenerate,
and implement the current limit $\Delta S=0.05\pm 0.10$\cite{PDG2016}.
To proceed, we assume that the mass squared differences, $|M_E^2-M_N^2|, |M_{T_0}^2-M_{T_-}^2|$, are at most $v^2$ (see Sec. 2). It is easy to generalize to other values.
We define two variables, $x_E\equiv (M_N/M_E)^2$ and $x_T=(M_{T_0}/M_{T_-})^2$, for the discussion.
In terms of these variables
\beqa
\Delta T&=& \frac{1}{8\pi s_W^2 M_W^2}\left[ 3 M_E^2 I_2(x_E)+ M_{T_-}^2 I_2(x_T)-16\pi^2 v_t^2\right]\,,\\
\Delta S&=&\frac{1}{\pi}(1+\ln x_E+\frac{1}{3}\ln x_T)\,,
\eeqa
where
\beq
I_2(x)=1+x- \frac{ 2x \ln x }{x-1}\,.
\eeq
The function $I_2(x=1)=0$ and it is monotonically increasing when $x$ goes to zero, $I_2(0)=1$.
The $2\sigma$ range of $\Delta S$, $-0.15< \Delta S<0.25$, amounts to
\beq
0.3317< (x_E^3 x_T)^{1/4}<0.8513\,.
\label{eq:Del_S_bound}
\eeq
Apparently, $x<1$ is preferred, and one needs either $M_N<M_E$, $M_{T_0} <M_{T_-}$ or both to satisfy the requirement form $\Delta S$.
Since we assume the mass squared difference is at most $v^2$. In the cases of largest mass squared splitting, $M_E, M_{T_-}$ can be related to $x$'s as
\beq
M_E^2= \frac{v^2}{1-x_E}\,,\;M_{T_-}^2= \frac{v^2}{1-x_T}\,.
\label{eq:TS_x_M}
\eeq
We consider two simple cases: $x_E=1$, and $x_T=1$.
For $x_T=1$, the direct search of charged Higgs sets a bound $M_{T_-}(=M_{T_0})>80$GeV\cite{PDG2016}.
$\Delta S$ requires that $0.2296<x_E<0.8069$. From $\Delta T$, one has
\beq
-0.16<\frac{1}{8\pi s_W^2 M_W^2}\left[ 3 M_E^2 I_2(x_E)-16\pi^2 v_t^2\right] <0.32\,.
\eeq
The largest $\Delta T_F$ comes from the smallest $x_E$, namely, the largest mass squared difference.
By using Eq.(\ref{eq:TS_x_M}), one obtains $ v_t<23.75$GeV, with $M_E=280.3$GeV, and $M_N=134.2$GeV.

By the same token, when $x_E=1$, one has $0.0121<x_T<0.5253$, $v_t < 19.72$GeV, with $M_{T_-}=247.5$ GeV, and $M_{T_0}=27.2$ GeV.
Since $T$ does not carry lepton number, it interacts with SM fermions through the mixing $U_h^{12}$ and $b\bar{b}$ will be the dominant decay channel.
However $t_0$ has the SM gauge interaction, see Table \ref{tab:TVV_vertex},
and the process $e^+e^-\ra Z^*\ra Z b\bar{b}$ can go with an effective mixing squared $\simeq (4v_t/v)^2|U_h^{12}|^2<0.013$.
The effective mixing squared agrees with the bound, $\lesssim 0.02$,  from the direct search of neutral scalar at LEP2 for this mass\cite{LEP2}.

 A full analysis yields an upper bound
\beq
v_t< 24.08 \; \mbox{GeV}\,,
\eeq
which corresponds to $x_E=x_T=0.3317$, $M_E=M_{T_-}=300.9$ GeV, $M_N=M_{T_0}=173.3$GeV.
The solution agrees with the current direct search bounds on the masses of charged heavy lepton,  $\gtrsim 100$GeV\cite{HeavyLatLEP}, and Higgs,  $\gtrsim 80$ GeV\cite{PDG2016}.
Moreover, $M_N$ and $M_{T_0}$ are larger than the LEP2 bound from the $Z$ decay and the direct search for the neutral Higgs.
Comparing to the previous bound, Eq.(\ref{eq:vtbound_no_loop}), where loop contributions are not included, the upper bound for $v_t$ is pushed up by
 around factor of five.

This much has been said about the upper bound on $v_t$. We should remark that the oblique parameters do not impose any lower bound on $v_t$.
For example, even $v_t=0$, all requirements from $\Delta T$ and $\Delta S$ also that the mass squared differences are less than $v^2$ can be met when $x_E=x_T=0.8513$, $M_E=M_{T_-}=613.7$GeV, and $M_N=M_{T_0}=566.2$GeV.

\section{Higgs to $2\gm$}
New electrically charged degrees of freedom which couple to $h_{SM}$ modify the SM Higgs di-photon decay width.
In addition to the new charged leptons introduced for anomaly cancelation, which have been studied in \cite{Chang:2018vdd},
the charged components of the triplet also contribute.
For $h_{SM}$ di-photon decay, the relevant Lagrangian are the  $\lambda_{1,6}$ terms,
\beq
\simeq v(\lambda_1+\lambda_6/2 )h_{SM} T_+T_- +  v(\lambda_1+\lambda_6 )h_{SM} T_{++}T_{--}\,,
\eeq
 assuming that $|U_h^{11}|\sim 1.0$.
Although the $\lambda_{4,t}$ terms also contribute to $h_{SM}T_+ T_-$ and $h_{SM}T_{++} T_{--}$ vertices, their strengthes are doubly suppressed by $v_t$ and $U_h^{12}$, and thus can be ignored.
The di-photon decay width is thus
\beqa
\Gamma(H\ra \gamma\gamma)&=& {G_F \alpha^2 M_H^3 \over 128\sqrt{2} \pi^3}
\left| F_1(\tau_W) +\frac{4}{3}F_{1/2}(\tau_t) +\sum_{i=1}^6  y_{E_i}\frac{2 M_W}{g_2 M_{E_i}}F_{1/2}(\tau_{E_i})\right.\nonr\\
&&\left.+ (\lambda_1+\lambda_6/2 ) \frac{v^2}{2 M_{T_+}^2}F_0(\tau_{T_+})+ 4 (\lambda_1+\lambda_6 ) \frac{v^2}{2 M_{T_{++}}^2}F_0(\tau_{T_{++}})
\right|^2\,,
\eeqa
where $\tau_i\equiv (m_H/2 m_i)^2$, and all the loop functions can be found in \cite{tome}.
For the exotic leptons,   the Yukawa couplings are parameterized as ${\cal L}\supset -  y_{E_i} \bar{E}_i E_i h_{SM}$ in the mass basis.
Assuming that $T_-$ and $T_{--}$ are degenerate, the width reads
\beqa
\Gamma(H\ra \gamma\gamma)&=& \left.{G_F \alpha^2 M_H^3 \over 128\sqrt{2} \pi^3}
\times \right| -8.324 + 1.834  \nonr\\
&& \left.  +\sum_{i=1}^6   0.32(3.64 )\times y_{E_i} + 0.051(0.203)\times (\lambda_1 +  0.9 \lambda_6)   \right|^2
\eeqa
 for  $M_{E_i}=1000(100)$GeV, and  $M_{T_-}=1.0(0.5)$TeV.
The first two numbers are the dominate SM contributions from $W^\pm$ and top quark, respectively.
 The dominant SM Higgs production channel at the LHC is through gluon fusion which is intact in this model.
Therefore, the signal strength of $pp\ra h\ra \gamma\gamma$ is
\beqa
\mu_{\gamma\gamma}&\simeq &\Gamma(H\ra \gamma\gamma)/\Gamma(H\ra \gamma\gamma)_{SM}\nonr\\
 &\sim& 1 - \sum_i(0.049-0.561)\times y_{E_i} -(0.0157-0.063)\times( \lambda_1+0.9\lambda_6)\,.
\eeqa
It is expected that $ |y_{E_i}| \sim m_l/v_h \ll 1$\cite{Chang:2018vdd}, and the charged leptons contribution can be ignored.
Comparing to the data $\mu_{\gamma\gamma}=1.18 (+0.17 -0.14)$ \cite {CMS18}, it is safe even $|\lambda_{1,6}|\sim{\cal O}(1)$.
This agrees with the general analysis given in \cite{CNWS}.

\section{Conclusions}

We have studied a novel neutrino mass generation mechanism in the recently proposed gauged lepton number model by us \cite{Chang:2018vdd}.
The model is free of anomalies by the addition of two sets of exotic chiral leptons for each generation.
The $U(1)_l$ gauge symmetry is spontaneously broken when a $l=1$ SM singlet, $\phi_1$, gets a VEV, $v_L$.
In addition, one $l=0$ SM triplet, $T$, is introduced for neutrino mass generation.
The triplet in this model differs from the well-studied $l=2$ triplet in the type-II see-saw model.
Since it carries no lepton number, the triplet does not couple to the SM leptons.
An immediate consequence is that there is no doubly charged triplet contribution to the neutrino-less double decays of nuclei which in our model is given mainly by the exchange of light neutrinos.
The VEV of the charge-neutral parts of $T$, $v_t$, and the SM Higgs $H$, $v\simeq 246$GeV, breaks the SM electroweak gauge symmetry
and the custodial symmetry.
With only two exotic scalars, $\phi_1$ and $T$, and no RH SM singlet neutrino, the resulting
neutrino mass is of the inverse see-saw type.
Since the phenomenology of the obligatory new gauge boson $Z_\ell$ and the exotic leptons have been studied in \cite{Chang:2018vdd}, we have focused on the
physics of neutrino mass and the new $l=0$ triplet in this work.

 We begin the discussion of the one-generation case since the physics is clear in this simple setting. Since the exotic leptons required for anomaly cancelations will in general mix with the SM leptons we require
  that the Yukawa couplings $f_{1,2}$ to be very small. This discussion is later extended to the realistic three-generation case, and we have carefully investigated the physics of active neutrino masses in this model.
The active neutrino masses are of the order of $v_t^3/v_L^2$ given by the dimension-six operator $O_6$. Since the electroweak precision requires a relatively small $v_t$, no further parameter fine-tuning is required other than taking $f_1\simeq 0$ mentioned before.
Both realistic NH and IH neutrino masses can be accommodated in this model. If assuming a democratic structure of the Yukawa couplings, it is more natural to get an NH pattern. For IH, it requires a more subtle Yukawa pattern and prefers to have the lightest neutrino mass $\gtrsim10^{-2}$ eV, which is promising for the neutrinoless double beta decays searches.

 It is worth noting that $O_6$ produces elements of the active neutrino mass matrix that is Majorana-like, i.e of the form
$\ovl{\nu^{ic}_{L}} \nu_{L}^j$ where $i,j$ are family indices. This is the same as $O_5$ would.
Thus, low energy neutrino measurements such as neutrinoless double beta decays of nuclei, tritium $\beta$ decays spectrum endpoint, and cosmological neutrino mass bounds cannot distinguish between $O_6$ or the Weinberg operator
as the origin of neutrino masses. In order to do that one needs to explore the TeV scale to discover whether there are new degrees of freedom. $O_5$ assumes that there are none whereas $O_6$ requires
new leptons below 10 TeV\footnote{ We have seen previously that the mass splitting $|M_E^2-M_N^2|\lesssim v^2$. Leptons with the mass around $10$TeV will give a splitting of $< 1$ GeV. This is much smaller than what we
have encountered and will require very delicate tuning of parameters.}. In addition, a detailed program searching for CLFV decays of muon and $\tau$ will also be useful since $O_5$ and $O_6$
have very different UV completions and thus will yield different results for these processes.

We have calculated the 1-loop triplet contributions to $a_\mu$ and the LFV processes $l_1\ra l_2 \gamma ( l_{1,2}=e,\mu,\tau )$.
$\Delta a_\mu$ is negative but negligible in this model. Thus, it cannot resolve the discrepancy
 between the data and the SM expectation \cite{PDGamu}. On the other hand, we have found an interesting connection between the neutrino masses and the LFV
branching ratios. Taking into account the current limit on $Br(\mu\ra e\gamma)<4.2\times 10^{-13}$, we have obtained an interesting lower bound on $v_t\gtrsim 0.1$ GeV.
Since $T$ does not couple to SM leptons, the LFV process $\mu\ra 3 e$ and the $\tau$ counterparts are mediated by the photon dipole transition
and thus predicted to be very small,  $Br(\mu\ra 3e)\lesssim 10^{-15}$.

The triplet gets a VEV so that the constraint from $\Delta T$ can be relaxed. We have carefully analyzed the limits from both $\Delta S$ and $\Delta T$ and arrived an upper bound for $v_t\lesssim 24.1$ GeV if assuming the mass squared differences among the isospin components of the triplet and the heavy leptons to be at most electroweak, $\lesssim v^2$. Combing with the neutrino masses and LFV bounds, we have $0.1 \lesssim v_t\lesssim 24.1$GeV in this model. The lower bound of $v_t$ also implies that $M_T\lesssim 8$TeV provided that $\kappa \simeq v$.

We have studied the decays of the triplet. For $T_\pm$ and $T_{\pm\pm}$, the dominant decay channel is into di-boson.
Depending on the scalar potential, the $T_0$ component of the triplet in general mixes with the SM Higgs doublet, although the mixing squared is limited to be smaller than $0.13$ at $2\sigma$ level\cite{LHC_mu}. However, even allowing for this mixing the dominant decay channel of $T_0$ is still the SM di-boson modes.
Due to their SM gauge interactions, the charged triplets can be pair produced via Drell-Yan processes at the LHC.
In addition to the SM gauge couplings, due to its mixing with the SM Higgs, the neutral triplet can be  singly produced via the gluon fusion. At LHC14, it is possible to probe  $t_0$ of mass up to 1TeV and $|U_h^{12}|^2\sim0.1$ with an integrated luminosity of $300 fb^{-1}$.
At the linear colliders, the signal of triplet pair production will be very clean once the center-of-mass energy is higher than the mass threshold.
For the mass range of triplet we are interested in, we have found that the bound from the current $h_{SM}\ra 2\gamma$ measurement is weak.

\begin{acknowledgments}
 WFC is supported by the Taiwan Minister of Science and Technology under
Grant No.106-2112-M-007-009-MY3 and No.105-2112-M-007-029.
TRIUMF receives federal funding via a contribution through the National Research Council of Canada.
\end{acknowledgments}

\bibliographystyle{JHEP}
\bibliography{Ref_Ul_2}

\end{document}